\documentclass[letterpaper,twocolumn,10pt]{article}

\usepackage[utf8]{inputenc}
\usepackage[T1]{fontenc}

\usepackage{usenix}
\usepackage{cite}
\usepackage{booktabs}
\usepackage{tabularx}
\usepackage{multirow}
\usepackage{makecell}

\usepackage{amsmath,amssymb}
\usepackage{mathrsfs}

\usepackage{graphicx}
\usepackage{booktabs}
\usepackage{array}
\usepackage{threeparttable}

\usepackage{xcolor}
\usepackage{xurl}
\usepackage{hyperref}
\newtheorem{definition}{Definition}

\usepackage{multicol}
\usepackage[ruled,linesnumbered]{algorithm2e}
\DontPrintSemicolon
\SetInd{0.5em}{1.2em}

\SetKw{Return}{return}
\SetKw{Assert}{assert}
\SetKwProg{On}{on}{}{end}
\newcommand{\on}[2]{\On{#1}{#2}}

\usepackage{balance}
\usepackage{xspace}
\usepackage{tikz}
\newcommand{\circled}[1]{%
  \tikz[baseline=(char.base)]{
    \node[shape=circle,draw,inner sep=0.7pt] (char) {\small #1};
  }%
}

 \newcommand{\eg}{\hbox{\emph{e.g.}}\xspace} \newcommand{\ie}{\hbox{\emph{i.e.}}\xspace}

\newcommand{\myparagraph}[1]{%
  \vspace{0.12cm}\noindent\textbf{#1.}\xspace
}
\newcommand{\Hash}{\mathcal{H}}
\SetKw{assert}{assert}

\usepackage{amssymb}

\begin{document}

\date{}

\title{\Large \bf A402: Binding Cryptocurrency Payments to Service Execution for \\ Agentic Commerce
}

\author{
{\rm Yue Li}\\
Peking University
\and
{\rm Lei Wang}\\
Shanghai Jiao Tong University
\and
{\rm Kaixuan Wang}\\
Shanghai Jiao Tong University
\and
{\rm Zhiqiang Yang}\\
Shanghai Jiao Tong University
\and
{\rm Ke Wang}\\
Zhongguancun Laboratory
\and
{\rm Zhi Guan}\\
Peking University
\and
{\rm Jianbo Gao}\\
Beijing Jiaotong University
}

\maketitle


\begin{abstract}
The rapid proliferation of autonomous AI agents is driving a shift toward agentic commerce, where agents are expected to autonomously invoke and pay for services. While blockchain-based payments offer a programmable foundation for such interactions, the recently proposed x402 standard fails to enforce end-to-end atomicity across service execution, payment, and result delivery.

In this paper, we present A402, a trust-minimized payment architecture
that securely binds cryptocurrency payments to service execution. A402 introduces Atomic Service Channels (ASCs), a new channel protocol that integrates service execution into payment channels, enabling real-time, high-frequency micropayments for agentic commerce. Within each ASC, A402 employs an atomic exchange protocol based on TEE-assisted adaptor signatures, ensuring that payments are finalized if and only if the requested service is correctly executed and the corresponding result is delivered. To further ensure privacy, A402 incorporates a TEE-based Liquidity Vault that privately manages the lifecycle of ASCs and aggregates their settlements into a single on-chain transaction, revealing only aggregated balances.
We implement A402 and evaluate it against x402 with integrations on both Bitcoin and Ethereum. Our results show that A402 delivers orders-of-magnitude performance and on-chain cost improvements over x402 while providing trust-minimized security guarantees. 
\end{abstract}

\section{Introduction}
\label{sec:intro}

The internet is undergoing a paradigm shift from a human-centric
\emph{content economy} to a machine-centric \emph{agentic economy},
driven by the rapid proliferation of autonomous AI
agents~\cite{rothschild2026agentic,yang2025agentic}.
Unlike traditional human users, these agents increasingly operate in a
fully automated manner: they dynamically discover, invoke, and compose
machine-callable services, such as cloud
functions, data oracles, and agent
skills~\cite{vaziry2025towards}.
In many cases, such interactions involve not only service execution
but also service payment: an agent may need to obtain access to \emph{a service from a previously unknown provider without any
pre-established account, contractual relationship, or prior trust}.
As such interactions become both \emph{autonomous} and
\emph{high-frequency}, they naturally give rise to an emerging form of
agentic commerce, in which individual service
executions may themselves become billable events.

This emerging agentic commerce exposes the lack of a suitable payment primitive.
Existing traditional financial infrastructures rely on identity-based authentication, such as Know-Your-Customer requirements, which assume that counterparties possess legal or physical identity and can establish persistent
relationships in advance.
These assumptions fit human-centric commerce, which typically adopts
coarse-grained monetization models such as subscriptions.
However, autonomous agents may lack legal identity, interact with
previously unknown service providers, and require low-latency, fine-grained
settlement for individual service execution.
In contrast, blockchain bind assets to cryptographic keys, decoupling asset ownership from physical identity.
This property makes blockchain-based payments a natural foundation for
agentic commerce, where agents can transact directly, securely, and
programmatically without manual
intermediation~\cite{nakamoto2008bitcoin,wood2014ethereum}.

\begin{figure}
    \centering
    \includegraphics[width=1\linewidth]{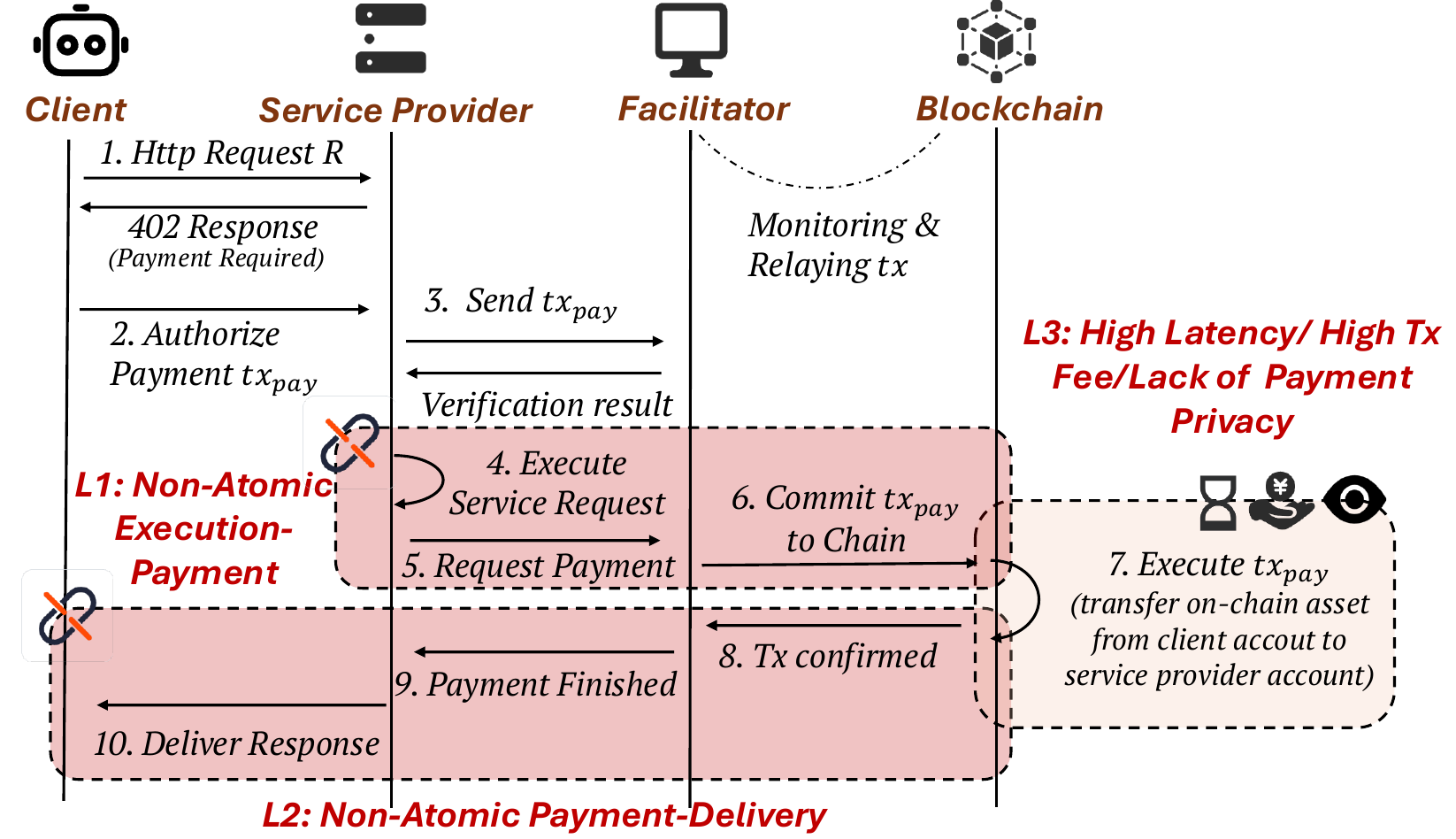}
    \caption{The workflow of x402 and its limitations}
    \label{fig:x402}
\end{figure}

To enable such agent payment, the industry has recently proposed x402, an emerging open standard that leverages the HTTP 402 "Payment Required" status code to enable clients to pay service providers with blockchain-based assets via a centralized facilitator~\cite{reppel2025x402}.
Within seven months after its May 2025 launch, x402 processed more than 100 million payments, reaching \$24 million across major facilitators like Coinbase, Cloudflare and Google~\cite{x402standard}. The workflow of x402 is shown in Figure~\ref{fig:x402}. When a client requests a service, the service provider responds with a 402 status and payment details (\emph{Step 1}). The client then signs the $tx_{pay}$ (\ie, payment transaction) and resubmits the request (\emph{Step 2}). The service provider forwards it to a centralized facilitator for validation (\emph{Step 3}). After that, the service provider executes the request (\emph{Step 4}) and then signals completion to the facilitator (\emph{Step 5}).
The facilitator then broadcasts the $tx_{pay}$ to the blockchain (\emph{Step 6}). Once $tx_{pay}$ is executed  and confirmed on-chain (\emph{Step 7}), the facilitator then notifies the service provider to deliver the response for the client (\emph{Steps 8-10}).

Despite its rapid adoption in practice, x402 fails to satisfy the security and performance requirements of high-frequency agentic commerce due to the following limitations: \textbf{L1.} service providers must execute requests optimistically before payment finalization, exposing them to non-payment risks. Moreover, facilitators broadcast payments without verifying that execution has been actually carried out. All these situations violate the atomic binding of execution and payment; 
\textbf{L2.} x402 enforces a payment-first workflow in which on-chain payment confirmation precedes result delivery. As a result, a malicious or failed service provider can withhold service responses after receiving payment. This violates fair exchange of payment and service response; 
\textbf{L3.} x402 suffers from scalability and privacy limitations: (i) the end-to-end latency is bounded by blockchain confirmation latency, rendering real-time high-frequency interactions impractical; (ii) high transaction fees render fine-grained micropayments economically infeasible; (iii) all interactions are recorded on-chain, compromising business privacy.

In this paper, we introduce A402, a secure agent payment architecture
that \emph{binds cryptocurrency payments to service execution},
enabling seamless agentic commerce while overcoming the fundamental
limitations of x402.
The core of A402 is a new channel protocol called Atomic Service Channel (ASC), which offloads on-chain payments into off-chain channels and embeds service execution into the payment channel’s state, allowing for low-latency micropayments in real-time interactions at a low cost (\textbf{cf. L3(i)(ii)}). Within the ASC, we design an atomic exchange protocol that combines TEE-based enforcement with adaptor signatures, which guarantees that a payment is finalized if and only if the request is correctly executed and the corresponding result is delivered, thereby enforcing the atomicity across execution, payment, and delivery (\textbf{cf. L1, L2}). 
To further address privacy concerns, we introduce a TEE-based Liquidity Vault that privately manages the creation and settlement of ASCs. By maintaining ASC state off-chain, the liquidity vault aggregates multiple ASC settlements into a single on-chain transaction that exposes only aggregated balances (\textbf{cf. L3(iii)}). 

Overall, the design of A402 makes the following contributions:



\begin{itemize}
    \item  We propose \textbf{Atomic Service Channels (ASCs)}, a new channel protocol that binds off-chain service execution to blockchain payments, enabling real-time, high-frequency micropayments.



    \item We design an \textbf{Atomic Exchange Protocol} using TEE-assisted adaptor signatures. This ensures end-to-end atomicity by linking payment finalization to the verifiable execution and delivery of service results.

    \item We propose a \textbf{TEE-based Liquidity Vault} to manage the ASC lifecycle privately off-chain. By aggregating multiple ASCs into a single on-chain settlement, it preserves privacy while ensuring settlement integrity.


\item  We implement A402 on both Bitcoin and Ethereum. Our prototype achieves a peak throughput of 2,875 RPS with sub-second latency ($\sim$350\,ms), representing orders-of-magnitude improvements in performance over x402 while enabling \emph{horizontal scalability}. Furthermore, A402 reduces on-chain cost from $O(n)$ (x402) to constant $O(1)$, providing a cost-effective and trust-minimized solution for M2M commerce.

\end{itemize}




\section{Preliminary}
\label{sec:background}

\subsection{Payment Channel and its Limitation}

\begin{figure}
    \centering
    \includegraphics[width=.8\linewidth]{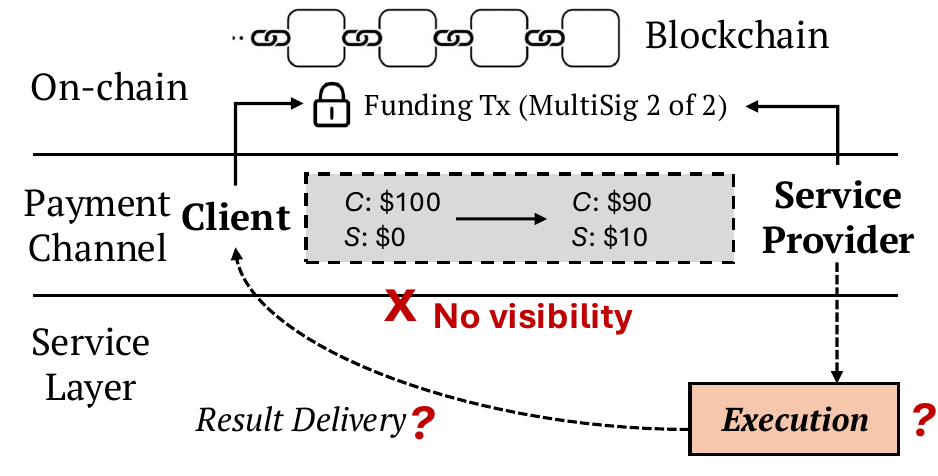}
    \caption{The workflow of payment channel and its limitations}
    \label{fig:paymentchannel}
\end{figure}

Payment channels enable two parties, a client ($C$) and a service provider ($S$), to process payments off-chain while relying on the blockchain merely for channel creation and final settlement (\ie, closure)~\cite{poon2016bitcoin}. Specifically, it manages intermediate state updates (\eg, assets updates) off-chain in order to 
significantly reduce on-chain overhead, and preserves the payment security guarantee since the channel settlement is enforced on-chain. As shown in Figure \ref{fig:paymentchannel}, we model a payment channel $\Pi_{PC}$ as a tuple consisting of three interfaces: 
\begin{itemize}

\item \emph{Channel Creation:} $\mathtt{Open}(v_C, v_S) \to (tx_{create}, \mathbf{B}_0)$. To establish the channel, $C$ and $S$ agree on their initial asset deposits, $v_C$ and $v_S$. The interface generates a transaction $tx_{create}$ which locks the total assets $(v_C + v_S)$ into a 2-out-of-2 multisig address on the blockchain. For example, as illustrated in Figure \ref{fig:paymentchannel}, the channel is initialized that the $C$ holds \$100 and the $S$ holds \$0.
Simultaneously, these two parties establish an initial off-chain state $\mathbf{B}_0$ reflecting this specific asset distribution.

\item \emph{Payment / State Transition}: $\mathtt{Pay}(\mathbf{B}_i, \delta) \to \mathbf{B}_{i+1}$.
To transfer asset $\delta$ from  $C$ to $S$, the parties negotiate a new state $\mathbf{B}_{i+1}$. This update is valid if and only if both parties exchange valid signatures over the new balance: $\mathbf{B}_{i+1} \leftarrow \mathtt{Sign}(\mathbf{B}_{i+1})$. This process occurs entirely off-chain and incurs no transaction fees. For example, the $C$ transfers $\$10$ to the service provider, updating the state from $(\$100, \$0)$ to $(\$90, \$10)$, as shown in Figure \ref{fig:paymentchannel}.

\item \emph{Channel Closure/ Settlement}: $\mathtt{Close}(\mathbf{B}_{final}) \to tx_{close}$.
To close the channel, either party can submit the latest valid state $\mathbf{B}_{final}$ (and its associated signatures) to the blockchain. The blockchain verifies the signatures against the $tx_{create}$ and distributes the locked assets back to $C$ and $S$ according to $\mathbf{B}_{final}$. Note that  $tx_{close}$ also initiates an on-chain \emph{dispute window} during which the counterparty ($C$ or $S$), or an optional \emph{watchtower}, can challenge the submitted $\mathbf{B}_{final}$ by providing a more recent state $\mathbf{B}_{new}$ with valid signatures. After the dispute window expires, the locked assets will be distributed on chain according to the latest valid $\mathbf{B}$.

\end{itemize}

\myparagraph{Lack of Execution–Payment Binding} Despite their efficiency, payment channels suffer from a fundamental limitation to be deployed in agentic commerce: they track only asset balances and are oblivious to service execution and result delivery. As shown in Figure \ref{fig:paymentchannel}, 
the channel state transitions are restricted to asset updates and provide no mechanism to represent, verify, or enforce off-chain service execution. As a result, payment channels cannot guarantee atomicity between service execution, payment, and result delivery.
This limitation causes two fairness violations. 
First, if $C$ co-signs an updated balance state $\mathbf{B}_{i+1}$ before receiving the service result, a malicious $S$ can immediately invoke $\mathtt{Close}(\mathbf{B}_{i+1})$ to finalize the payment on-chain while aborting execution.
Conversely, if $S$ executes the request first, the $C$ may refuse to co-sign the corresponding asset update, obtaining the service without paying (\ie, free-riding).

In both cases, the blockchain cannot enforce fairness because channel settlement lacks verifiable evidence of off-chain execution, fundamentally limiting the use of payment channels for agentic commerce. This limitation extends to protocols built atop payment channels. For instance, L402 leverages Lightning payment channels for HTTP 402 micropayments. However, after payment, the client in L402 has no cryptographic guarantee on the correctness or delivery of the service result.


\subsection{Adaptor Signature}

Adaptor signatures extend standard digital signature schemes (\eg, Schnorr digital signature~\cite{wuille2020bip340}) to enable atomic exchange between a valid signature and a cryptographic secret $t$~\cite{gerhart2024foundations}. Intuitively, an adaptor signature allows a signer to produce a \emph{pre-signature} $\hat{\sigma}$ with respect to a statement $T = t\cdot G$, which can be completed into a valid signature $\sigma$ if and only if a corresponding secret witness $t$ is revealed. Here, $t \in \mathbb{Z}_q$ is a secret sampled from the finite field of order $q$, and $G$ is a generator of the elliptic curve group. 
Overall, an adaptor signature scheme $\Pi_{AS}$ over a hard relation $R$ consists of the following four interfaces:
\begin{itemize}

\item $\hat{\sigma} \leftarrow \mathtt{pSign}(sk, m, T)$: The signer generates a pre-signature $\hat{\sigma}$ on message $m$ locked to statement $T$. This signature is invalid under the standard verification logic but cryptographically binds to $T$.

\item  $0/1 \leftarrow \mathtt{pVerify}(pk, m, T, \hat{\sigma})$: A verifier checks that $\hat{\sigma}$ is a well-formed pre-signature with respect to the statement $T$. 
If verification succeeds, it guarantees that \emph{given} the corresponding witness $t$, 
the pre-signature $\hat{\sigma}$ can be transformed into a valid signature.


\item $\sigma \leftarrow \mathtt{Adapt}(\hat{\sigma}, t)$: Given the witness $t$ for the statement $T$, any party can transform the pre-signature $\hat{\sigma}$ into a valid signature $\sigma$.


\item $t \leftarrow \mathtt{Extract}(\sigma, \hat{\sigma}, T)$: Given the pre-signature $\hat{\sigma}$ and the completed signature $\sigma$, any observer can efficiently extract the witness $t$.
\end{itemize}

\subsection{Trusted Execution Environment}
Trusted Execution Environments (TEEs) provide hardware-enforced isolation that protects the confidentiality and integrity of code and data even in the presence of a compromised operating system or hypervisor. Early TEE designs, such as ARM TrustZone~\cite{pinto2019demystifying} and Intel SGX~\cite{costan2016intel}, focused on enclave-level isolation within a single process. In contrast, modern TEE platforms, including Intel TDX\cite{cheng2024intel} and AMD SEV-SNP\cite{amd2023sev_snp_spec}, support \emph{Confidential Virtual Machines (CVMs)}, extending hardware-backed protection to entire virtual machines and shielding guest execution from malicious or compromised hypervisors. These TEEs further provide remote attestation, which allows a remote party to verify that a specific software stack is executing inside a genuine TEE instance.
Formally, a remote attestation scheme exposes the following interfaces:

\begin{itemize}
\item $\sigma_{att} \leftarrow \mathtt{Attest}(m)$: A TEE produces an attestation report $\sigma_{att}$ that cryptographically binds a message $m$ to a hardware-verified measurement of the TEE-protected execution environment.

\item $\{0,1\} \leftarrow \mathtt{VerifyAtt}(\sigma_{att}, m)$: A remote verifier checks that $\sigma_{att}$ was generated by a genuine TEE instance running the expected software stack, thereby establishing trust in the integrity of $m$.
\end{itemize}

\section{Overview}
\label{sec:overview}
$\textsc{A402}$ is a trust-minimized protocol designed to facilitate secure Web 3.0 payments for Web 2.0 services. At its core, $\textsc{A402}$ introduces Atomic Service Channels and leverages TEE-assisted adaptor signatures to enforce the \emph{Exec–Pay–Deliver atomicity} invariant for each service request.


\subsection{The Workflow of A402}

As illustrated in Figure~\ref{fig:architecture}, the A402 architecture consists of four entities:
\begin{itemize}
\item \textbf{Client ($C$).} 
The $C$ represents an autonomous agent requesting services off-chain. 
Being resource-constrained and potentially offline, the $C$ delegates channel management and service invocation with payment to the vault, which returns the service results.



\item \textbf{Vault ($\mathcal{U}$)}: 
The vault ($\mathcal{U}$) is a TEE-protected service responsible for managing Atomic Service Channels (ASCs) between clients and service providers. 
Acting as a delegated ASC manager, the $\mathcal{U}$ maintains ASC state, forwards service requests and results, and enforces ASC protocol logic within a TEE instance. 
To improve scalability and availability, the $\mathcal{U}$ may be deployed as multiple TEE-backed replicas that execute identical protocol logic and maintain consistent ASC state. 
In the \emph{liquidity vault mode}, the $\mathcal{U}$ can accept asset deposits from $C$ to pre-fund ASCs, enabling private ASC creation and closure as well as privacy-preserving settlement.


\item \textbf{Service Provider ($S$)}: 
The service provider executes service requests forwarded by the $\mathcal{U}$ via ASCs. A402 employs the TEE-assisted adaptor signature scheme to ensure that $S$ can only finalize the payment if and only if the service request is executed and the result is delivered.

\begin{figure}[t]
    \centering
    \includegraphics[width=\linewidth]{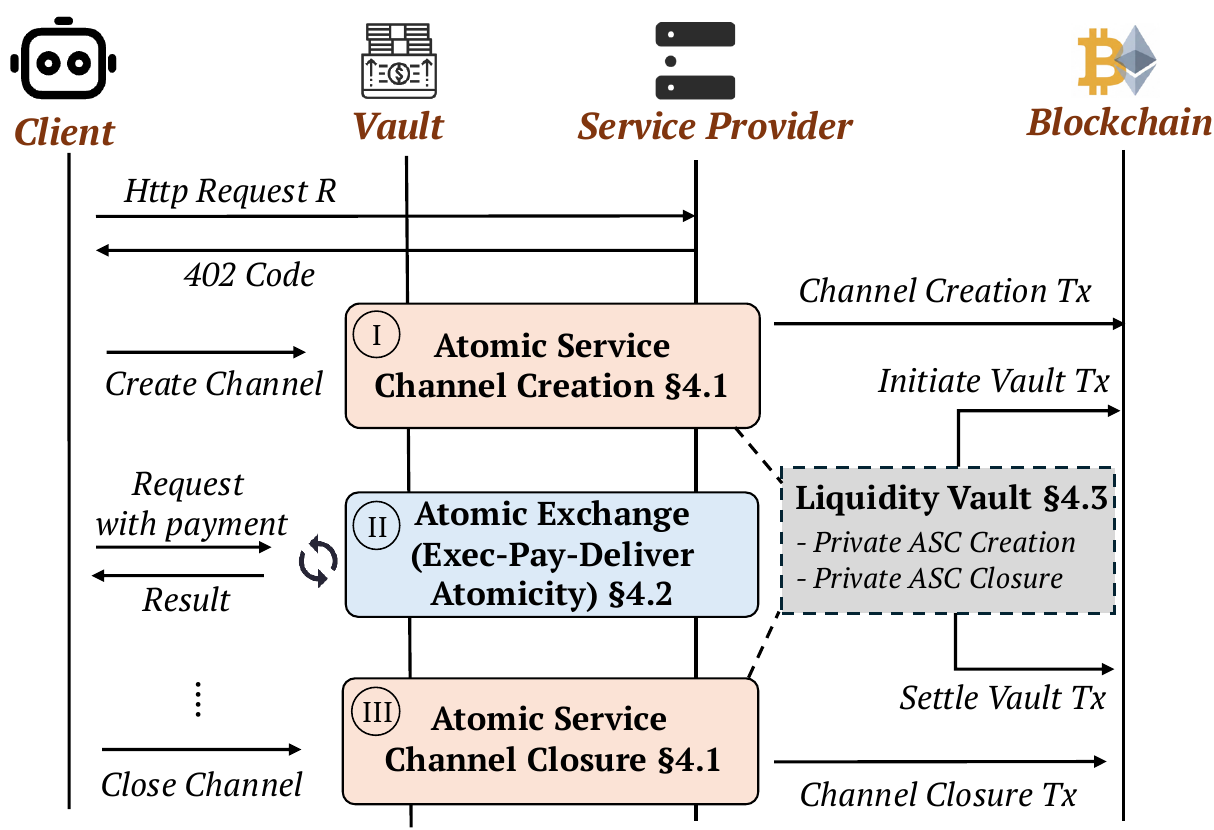} 
    \caption{A402 Overview.}
    \label{fig:architecture}
\end{figure}


\item \textbf{Blockchain / On-Chain Program ($\mathcal{L}$).}
A402 relies on the underlying blockchain as a public settlement and dispute resolution layer. The on-chain logic is implemented either as a smart contract on account-based chain (\eg, Ethereum) or as UTXO scripts on Bitcoin. It escrows ASC assets at creation and enforces conditional settlement at closure. 
In the \emph{liquidity vault mode}, it additionally governs the initialization and settlement of the liquidity vault.

\end{itemize}

The A402 protocol operates over three phases: ASC creation, atomic exchange, and ASC closure. In addition, A402 optionally supports a \emph{privacy-preserving liquidity vault} mode that is orthogonal to the ASC lifecycle.

\myparagraph{Phase I: Atomic Service Channel Creation}
The lifecycle begins when a client $C$ requests a service and receives a 402 Payment Required response. To proceed, $C$ delegates the session to the $\mathcal{U}$, which interacts with the on-chain program $\mathcal{L}$ to establish an ASC between $C$ and $S$. This phase escrows the client-approved assets on-chain and initializes the ASC state for subsequent off-chain interaction.

\myparagraph{Phase II: Atomic Exchange}
Once the ASC is established, A402 enters a high-frequency off-chain interaction phase. The $\mathcal{U}$ receives service requests from $C$ and forwards them to $S$ together with conditional payments. Using the \emph{atomic exchange protocol}, the $S$ executes the request and returns the corresponding result, while payment is finalized if and only if execution and delivery are successfully completed. This phase supports multiple service invocations without additional on-chain interaction.

\myparagraph{Phase III: Atomic Service Channel Closure}
Upon session termination, either $C$ or $S$ may initiate channel closure based on the latest ASC state. In the standard mode, closure is realized by submitting
an on-chain transaction to $\mathcal{L}$. The $\mathcal{L}$ verifies the submitted ASC state and, when necessary, opens a dispute window during which the counterparty may present a more recent valid ASC state.
After that, $\mathcal{L}$ distributes the escrowed assets according to the latest ASC state. 

\myparagraph{Privacy-Preserving Liquidity Vault}
In addition to on-chain ASC creation and closure, A402 supports private ASCs via a TEE-based liquidity vault. In this mode, $C$ deposits assets into $\mathcal{U}$ via $\mathcal{L}$ in advance. ASC creation and closure are handled off-chain, while the settlement of multiple ASCs is \emph{aggregated} into a single on-chain transaction at the vault level, reducing blockchain-observable leakage about individual ASCs (\eg, the client-service provider interaction graph and
per-channel capacity).



\subsection{Threat Model}
\label{subsec:threat}

We describe the trust assumptions and adversarial capabilities considered in A402.

\myparagraph{Trusted Hardware Assumptions}
A402 adopts a trust-minimized model based on the hardware-enforced guarantees of Trusted Execution Environments (TEEs). We assume the TEE provides confidentiality and integrity for protected code and data even in the presence of a compromised host operating system or hypervisor in line with other TEE-assisted proposals~\cite{das2019fastkitten,kuefner2021bitcontracts,wust2020ace,li2025ethercloak,wen2025mercury,bentov2019tesseract,frassetto2023pose}. In particular, an adversary controlling the host cannot read or modify TEE-protected memory, extract sealed cryptographic secrets, or cause the TEE to deviate from its attested execution logic. We assume the hardware vendor’s remote attestation infrastructure is trustworthy and uncompromised. Consistent with prior TEE-assisted proposals~\cite{das2019fastkitten,kuefner2021bitcontracts,wust2020ace,li2025ethercloak,wen2025mercury,bentov2019tesseract,frassetto2023pose}, we treat side-channel leakage and other TEE-specific vulnerabilities as out of scope.

\myparagraph{Adversarial Parties}
We consider a powerful adversary that may corrupt any subset of protocol participants outside the TEE trust boundary: \emph{(i) Malicious Client ($\mathcal{C}_{adv}$).}
A malicious client may attempt to obtain service results without payment (free-riding) or double-spend assets across multiple atomic service channels.
\emph{(ii) Malicious Vault ($\mathcal{U}_{adv}$).}
A malicious vault may attempt to observe, modify, or censor ASC workflow. Nevertheless, $\mathcal{U}_{adv}$ cannot tamper with ASC state, or violate code enforced inside the TEE.
\emph{(iii) Malicious Service Provider ($\mathcal{S}_{adv}$).}
A malicious service provider may drop or delay messages, abort service execution, or attempt to claim payment without executing the requested service. However, due to TEE enforcement, it cannot forge execution results, alter the attested execution logic, or extract secret keys. 
Beyond A402 participants, we consider a passive adversary monitoring on-chain transactions to construct the ASC interaction graph. The adversary aims to deanonymize client-service provider relationships and infer ASC capacity.


\myparagraph{Network Adversary}
We assume a fully asynchronous network controlled by the adversary. The adversary may eavesdrop, drop, replay, inject, or arbitrarily delay messages exchanged among $C$, $\mathcal{U}$, $S$, and the blockchain. We assume reliable eventual message delivery for liveness, but make no synchrony assumptions for safety.

\myparagraph{Blockchain Assumptions}
A402 anchors its safety on the underlying blockchain. We assume the blockchain provides persistence (safety) and eventual transaction inclusion (liveness) under standard consensus assumptions\cite{garay2024bitcoin,pass2017analysis}. Specifically, we assume that the on-chain program executes correctly and that block producers will eventually include valid transactions.

\subsection{Design Goal}

Based on the threat model described above, we derive the following desirable security properties for A402:

\myparagraph{Trust-Minimized Asset Security}
A402 must ensure that clients and service providers retain sovereign control over their assets under a trust-minimized model. Asset safety must not depend on the liveness, availability, or honesty of the vault. Even in the event of complete vault censorship, compromise, or service shutdown, A402 must provide an on-chain escape mechanism that allows parties to reclaim their assets and prevent indefinite asset locking.


\myparagraph{Exec-Pay-Deliver Atomicity} 
A402 must enforce end-to-end atomicity across service execution, payment, and result delivery. Payment may be finalized if and only if the request is correctly executed, and the corresponding result must be delivered once payment is finalized. No party should be able to obtain payment without execution, or obtain execution results without payment.


\myparagraph{Unlinkability} In privacy-preserving mode, A402 should decouple the off-chain ASC from the on-chain transactions.
An adversary observing the blockchain should be unable to link a specific on-chain transaction to a particular ASC, nor infer fine-grained service usages such as the frequency, volume, or interactions between clients and service providers.


Furthermore, we derive the following desirable functional properties for A402:

\myparagraph{Low-Latency} 
A402 must support low-latency, interactive services suitable for autonomous machines (\eg, AI agents).
During steady-state operation with honest participants (\ie, assuming the absence of disputes), the end-to-end latency of a service request should be dominated by network communication and service execution, without waiting for blockchain confirmation.

\myparagraph{Cost Efficiency} A402 must minimize on-chain interactions to reduce transaction fees and enable economically viable micropayments for agentic commerce.
Specifically, the on-chain cost should remain constant $O(1)$ regardless of the number of service requests.

\section{Design}
\label{sec:design}


In this section, we present the detailed design of A402, focusing on three core modules: Atomic Service Channel (Section~\ref{sec:design_exec_pay}), the Atomic Exchange Protocol (Section \ref{sec:design_pay_deliver}), and the Privacy-Preserving Liquidity Vault (Section \ref{sec:design_privacy}).

\subsection{Atomic Service Channel}
\label{sec:design_exec_pay}

\begin{figure}
    \centering
    \includegraphics[width=\linewidth]{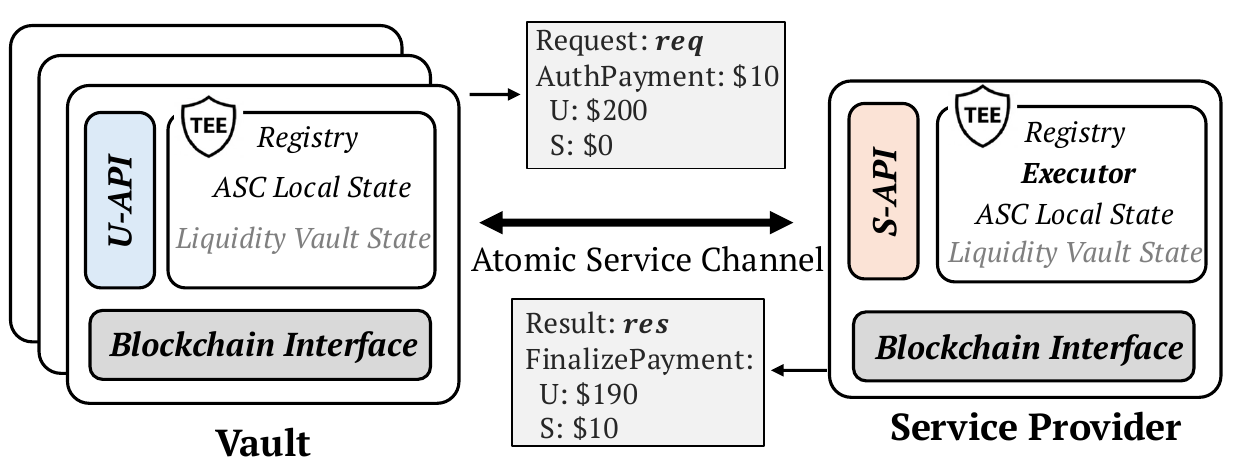}
    \caption{An Atomic Service Channel in A402. The details of U-API, S-API, and blockchain interfaces are described in the Table~\ref{tab:api_summary} of Appendix~\ref{appendix:interface}.}
    \label{fig:ASC}
\end{figure}

Traditional payment channels only capture asset transfers and are agnostic to off-chain service execution, making them unsuitable for agentic commerce. Atomic Service Channels extend this protocol by explicitly binding each payment to a service execution and its result delivery.
As illustrated in Figure \ref{fig:ASC}, instead of exchanging asset updates alone, the vault forwards a service request $req$ together with a conditional payment to the service provider. The service provider executes the $req$, finalizes the payment and returns the result $res$. 
We now present the formal definition of the \emph{Atomic Service Channel (ASC)}.

\begin{definition}[Atomic Service Channel (ASC)]
An \emph{Atomic Service Channel} with respect to a blockchain $\mathcal{L}$ is defined as a tuple
$(C, S, \Gamma)$, where $C$ is a client requesting off-chain services, $S$ is a service provider, and $\Gamma$ denotes the current ASC state of the channel.
The
$\Gamma = ( \mathbf{B}, \varsigma)$,
where $\mathbf{B} =\langle b_c, b_s\rangle$ represents the asset distribution in the channel, with $b_c$ held by the client and $b_s$ held by the service provider.
And $\varsigma$ denotes the execution status of channel.

An ASC supports the following three operations:
\begin{itemize}

\item \textbf{Channel Creation.}
$\mathtt{openASC}(C, S, v) \rightarrow \{\Gamma_0, tx_{create}\}.$
Given a client $C$, a service provider $S$, and an initial deposit $v$ provided by $C$,
the interface generates an on-chain locking transaction
$tx_{create}$ on $\mathcal{L}$ that locks $v$ assets under the joint control of $C$ and $S$.
It initializes the ASC state as $\Gamma_0 = ( \mathbf{B}_0, \varsigma_0)$
where $\mathbf{B}_0 = \langle v, 0\rangle$ and $\varsigma_0 = \texttt{Open}$ denotes that no service request is currently being processed.


\item \textbf{Atomic Exchange.}
$\mathtt{execPayDeliver}(\Gamma_k, req, \delta) \rightarrow \{\Gamma_{k+1}, res, 0/1\}.$
On input the current ASC state $\Gamma_k$, a service request $req$, and a payment amount $\delta$,
this interface atomically executes the requested service, processes the payment, and delivers the execution result.
If the execution completes successfully and the payment is processed, the ASC state is updated to
$\Gamma_{k+1} = ( \langle b_{C} - \delta, b_{S} + \delta\rangle, \texttt{Open})$,
and the execution result $res$ is delivered to the client and returned together with output $1$.
Otherwise, neither the state transition nor the result delivery occurs, and the operation returns $0$.

\item \textbf{Channel Closure / Settlement.}
$\mathtt{closeASC}(\Gamma_{final}) \rightarrow tx_{close}.$
On input the final ASC state $\Gamma_{final} = ( \mathbf{B}_{final}, \varsigma)$, either party may initiate channel closure by submitting $tx_{close}$ with $\Gamma_{final}$ to the blockchain $\mathcal{L}$. Upon invocation of $\mathtt{closeASC}$, the status $\varsigma$ transitions to \texttt{Closed}, indicating that the ASC is no longer active and will accept no further service requests.
The submitted transaction $tx_{close}$ initiates a dispute window, during which the counterparty may challenge the closure by presenting a more recent valid channel state. If no valid challenge is raised within the dispute window, the protocol finalizes settlement by distributing the locked assets according to $\mathbf{B}_{final}$ on chain.

\end{itemize}
\end{definition}

We now describe how ASCs are instantiated within the A402 architecture by mapping these ASC interfaces to the vault $\mathcal{U}$, the service provider $S$, and the on-chain program $\mathcal{L}$.

\myparagraph{TEE Registration}
A402 relies on TEEs to enforce protocol correctness without trusting the host operating system. Before participating in any ASC operation, both $\mathcal{U}$ and $S$ must register their TEEs.
As shown in Algorithm \ref{alg:asc} (lines 12–15), a TEE registers by submitting its public key $pk$, code hash $h_{code}$ and a remote attestation report $att$. The registration succeeds only if the attestation verifies correctly, binding $pk$ to an approved TEE instance measurement. Both the vault $\mathcal{U}$ and the service provider $S$ maintain a local registry (lines~2--5).
The service provider registry $\texttt{Reg}_S$ maps each service provider identifier $sid$ to its public key and code hash $(pk_S, h_{code,S})$, while the vault registry $\texttt{Reg}_U$ maps each vault identifier $uid$ to $(pk_U, h_{code,U})$.
After successful registration, $\mathcal{U}$ and $S$ can authenticate each other and establish a secure channel via an authenticated Diffie–Hellman key exchange.
The derived channel key $ch$ (line~5) is used to protect all subsequent ASC messages, providing confidentiality and integrity against a network adversary.

\myparagraph{ASC State and Lifecycle}
Each atomic service channel is represented by a local channel state
$\Gamma$ maintained by the vault $\mathcal{U}$ and  synchronized with the service provider $S$
(Algorithm~\ref{alg:asc}, lines~7--11).
Formally,
\[
\Gamma = (C, S, \mathbf{B}, \varsigma, k),
\]
where $C$ denotes the client, the $\varsigma$ denotes the channel status, and the version counter $k$ is incremented on each request to prevent replay. 
The $\mathbf{B}$ denotes the asset state and is defined as
\[
\mathbf{B} = \langle b^{C}_{free}, b^{C}_{locked}, b^{S} \rangle,
\]
where $b^{C}_{free}$ denotes the client’s available assets, $b^{C}_{locked}$ denotes assets temporarily locked for an in-flight request, and $b^{S}$ denotes the service provider’s accumulated revenue within the ASC. The lifecycle of an ASC is captured by the channel status $\varsigma$, which transitions through the following stages:

\begin{algorithm*}[t]
\caption{Atomic Service Channel}
\label{alg:asc}
\scriptsize
\begin{multicols}{3}

\textbf{A402 State:}\;

\vspace{0.4em}
\noindent
\textbf{(i) Registry:}\;
\quad $\mathtt{Reg}_S : sid \mapsto (pk_S, h_{code,S})$\;
\quad $\mathtt{Reg}_U : uid \mapsto (pk_U, h_{code,U})$\;
\quad $\mathtt{SC} : id \mapsto ch$

\vspace{0.4em}
\noindent
\textbf{(ii) ASC State:}\;
\quad $\mathtt{ASC} : cid \mapsto \Gamma_{cid}$\;

\vspace{0.2em}
\noindent
\quad $\Gamma_{cid} = (C, S, \mathbf{B}, \varsigma, k)$ \;
\quad $\mathbf{B} = \langle b^{C}_{free}, b^{C}_{locked}, b^{S}\rangle$ 

\quad  $\varsigma \in \{
    \texttt{OPEN},
    \texttt{LOCKED},
    \texttt{PENDING},
    \texttt{CLOSED}
\}$

\quad $k$ denotes a increasing state version counter.


\vspace{0.6em}

\on{$\mathtt{Register}(pk, h_{code}, att$)}{
 \assert $\mathtt{VerifyAtt}(att, pk, h_{code})$\;
    $\mathtt{Reg}_S/\mathtt{Reg}_U.\mathtt{add}(sid/uid)$\;
    
}
\vspace{0.6em}

\on{$C.\mathtt{reqOpenASC}(\mathcal{U}, S, v, \mathcal{L}, \pi)$}{
    $\mathcal{U}.\mathtt{initCommittee}(\pi)$\;
    \assert $\mathcal{U}.\mathtt{hasASC}(C,S, \mathtt{ASC})$\;
    \If{$\neg\mathcal{U}.\mathtt{hasSC}(S, SC)$}{
        $\mathcal{U}.\mathtt{establishSC}(S,SC)$\;
    }
    $ (cid, \mathbf{B}_{0}, tx_{create}) \gets \mathcal{U}.\mathtt{openASC}(C,S,v)$\;
    \assert $\mathcal{L}.\mathtt{exec}(tx_{create})$\;
    $ \mathtt{ASC}.\mathtt{add} (cid, \mathbf{B}_{0})$\;
    \Return $(cid, \mathbf{B}_{0})$;
}

\vspace{0.6em}

\columnbreak
\setcounter{AlgoLine}{26}
\on{$C/S.\mathtt{reqCloseASC}(cid)$}{
    $\Gamma \gets \mathcal{U}.\mathtt{getASC}(cid)$\;
    \assert $\Gamma.\varsigma \neq \texttt{CLOSED}$ \;
        \eIf{$\Gamma.\varsigma = \texttt{OPEN}$}
        {
           $tx_{close} \gets \mathcal{U}.\mathtt{closeASC} \;(\Gamma)$\;
        \assert $\mathcal{L}.\mathtt{exec}\;(tx_{close})$\;
         $\Gamma.\varsigma \leftarrow \texttt{CLOSED}$\;
        sync $\Gamma$ in both $\mathcal{U}$ and $S$
        }{
        $\mathcal{U}.\mathtt{waitUntil}(\Gamma.\varsigma = \texttt{OPEN})$\;
        retry\;
        }
        
}
\vspace{0.6em}

\on{$C.\mathtt{forceCloseASC}(cid, \Gamma)$}{

    $tx_{init} \gets C.\mathtt{initClose}(cid, \Gamma)$\;
        \assert $\mathcal{L}.\mathtt{exec}(tx_{init})$\;
        $\mathcal{L}.\mathtt{startDisputeTimer}(\Delta)$\;
    \If{$\Delta$ expired and no valid challenge from $S/\mathcal{U}$} 
    {
    $tx_{final} \gets C.\mathtt{finalClose}(tx_{init})$\;
        \assert $\mathcal{L}.\mathtt{exec}(tx_{final})$\;
        $\Gamma.\varsigma \leftarrow \texttt{CLOSED}$\;
     sync $\Gamma$ in both $\mathcal{U}$ and\ $S$\;
    }
        
}

\vspace{0.6em}

\columnbreak
\setcounter{AlgoLine}{50}

\on{$S.\mathtt{forceCloseASC}(cid, \sigma_S)$}{
    $\Gamma \gets \mathcal{S}.\mathtt{getASC}(cid)$\;
\assert $\Gamma.\sigma \neq \texttt{CLOSED}$\;
    \If{$\Gamma.\sigma = \texttt{OPEN}$}{
       
        $tx_{close} \gets \mathcal{C}.\mathtt{createClose}(\Gamma)$\;
        \assert $\mathcal{L}.\mathtt{exec}(tx_{close})$\;
        $\Gamma.\varsigma \gets \texttt{CLOSED}$\;
        sync $\Gamma$ in both $\mathcal{U}$ and $S$ \;
    }

    \If{$\Gamma.\varsigma = \texttt{PENDING}$}{
         $tx_{close}^{ASig} \gets \mathtt{closeWithASig}(\Gamma, \sigma_S)$\;
            \assert $\mathcal{L}.\mathtt{exec}(tx_{close}^{ASig})$\;
        $\Gamma.\varsigma \gets \texttt{CLOSED}$\;
        sync $\Gamma$ in both $\mathcal{U}$ and $S$\;
    }
}

\end{multicols}
\end{algorithm*}

\begin{itemize}
    \item \texttt{OPEN}: the channel is active and accepts new service requests;
    \item \texttt{LOCKED}: assets are escrowed for an in-flight request whose execution is ongoing;
    \item \texttt{PENDING}: execution is complete, and a conditional payment has been issued by $\mathcal{U}$; $\mathcal{U}$ awaits result revelation;
    \item \texttt{CLOSED}: the channel has been settled and cannot be reused.
\end{itemize}

\myparagraph{Atomic Service Channel Creation} 
ASC creation is initiated by the client via $\mathtt{reqOpenASC}$ (Algorithm~\ref{alg:asc}, lines 16–26).
Upon receiving the request, the $\mathcal{U}$ instantiates a committee according to the client-specified policy $\pi$ (\eg, $m$-out-of-$n$) and proceeds only after collecting sufficient authorizations from the committee members (line 17). The $\mathcal{U}$ then verifies that no active ASC already exists between the same client $C$ and service provider $S$, preventing duplicate channel creation (line 18). Next, $\mathcal{U}$ ensures the existence of a secure channel with $S$. If no such channel is available, the vault establishes one on demand (lines 19–21). 
The $\mathcal{U}$ then executes $\mathtt{openASC}$ to generate a channel identifier $cid$, an initial state $\Gamma_0$, and an on-chain creation transaction $tx_{create}$ (line 22). Once the $tx_{create}$ is confirmed on the underlying chain $\mathcal{L}$ (line 23), the ASC state is recorded locally and the channel transitions to the \texttt{OPEN} state, ready to process service requests (as described in Section~\ref{sec:design_pay_deliver}).

\myparagraph{Atomic Service Channel Closure}
ASC closure in A402 supports both cooperative and unilateral settlement, guaranteeing asset recovery under adversarial conditions and thereby ensuring
\emph{trust-minimized asset security}.

\textbf{\emph{Cooperative Closure.}}
Under standard cooperative conditions, either $C$ or $S$ may request ASC closure via the vault $\mathcal{U}$ (Algorithm~\ref{alg:asc}, lines~27–38). 
If the ASC is in the \texttt{OPEN} state, the $\mathcal{U}$ immediately settles it on-chain using the latest agreed asset state $\Gamma$ and transitions the ASC to \texttt{CLOSED} (lines 31-34). If an in-flight request exists (\ie, the ASC is in the \texttt{LOCKED} or \texttt{PENDING} state), settlement is deferred until the request completes, preventing partial execution from being settled (lines 36-37).

\textbf{\emph{Unilateral Client Closure.}} To ensure \emph{trust-minimized asset security}, A402 supports unilateral client
closure via $\mathtt{forceCloseASC}$ (lines~40-50), which bypasses the $\mathcal{U}$.
In this force-close path, the $C$ submits an initial close transaction $tx_{init}$ to the $\mathcal{L}$, containing the latest ASC state $\Gamma$ received from the $\mathcal{U}$ (lines~40–41). This transaction opens an on-chain dispute window of duration $\Delta$, during which the $S$ may challenge the closure by submitting a more recent valid ASC state (lines~42–43)\footnote{$S$ utilizes a so-called \emph{watchtower} as in payment channel to detect and respond to unilateral closure events within the challenge period}.
If no valid challenge is submitted before the dispute window expires, the $C$ finalizes settlement by submitting a transaction $tx_{final}$ (lines~45–47). Upon confirmation, the ASC transitions to the \texttt{CLOSED} state and assets are distributed according to the $\textbf{B}$ in the final state $\Gamma$. This mechanism guarantees asset recovery even under complete vault unavailability or service provider failure.

\textbf{\emph{Unilateral Service Provider Closure.}}
The service provider $S$ may also unilaterally close a ASC when the vault $\mathcal{U}$ is unavailable
(lines~51-66). 
If the ASC is in the \texttt{OPEN} state, the $S$ directly generates and submits a standard close transaction $tx_{close}$ using the latest state (lines~53–57). Once confirmed, the ASC transitions to the \texttt{CLOSED} state and settlement completes normally.

If the ASC is in the \texttt{PENDING} state, 
the $S$ generates the adaptor signature $\sigma_s$ for the current request
and submits a closing transaction $tx_{close}^{ASig}$ carrying the $\sigma_s$ (line 61-62). Upon confirmation, the $\sigma_s$ irrevocably reveals the execution witness $t$ on-chain, enabling the $S$ to claim its revenue while simultaneously allowing the vault $\mathcal{U}$ to extract $t$ (lines~64–66) (More details are provided in Section \ref{sec:design_pay_deliver}).

Note that the $S$ has no incentive to submit a stale ASC state, as doing so would only reduce its own revenue. Consequently, force closure via $S$ does not require a dispute window.

\begin{algorithm*}[t]
\caption{Atomic Exchange Protocol (Exec-Pay-Deliver Atomicity)}
\label{alg:atomicexchange}
\scriptsize
\begin{multicols}{3}

\noindent\textbf{Request State:}\;
\vspace{0.2em}
\noindent
\quad $\mathtt{Pend}:(cid,rid)\mapsto \mathcal{R}_{cid,rid}$\;
\quad $\mathcal{R}_{cid,rid}=(\delta,req,EncRes,\hat{\sigma}_S,T)$\;

\vspace{0.6em}
\on{$\mathcal{U}.\mathtt{sendRequest}(cid, req, \delta)$}{
    $(C,S,\mathbf{B},\varsigma,k) \leftarrow \Gamma_{cid}$\;
    \assert $\varsigma=\texttt{OPEN}\ \& \ b^{C}_{free} \ge \delta$\;

    $rid \leftarrow (cid,k)$\;
    $b^{C}_{free} \mathrel{-}= \delta;$ $b^{C}_{locked} \mathrel{+}= \delta$\;

    $\mathtt{Pend}[cid,rid] \leftarrow (\delta,req,\bot,\bot,\bot)$\;
    $\varsigma \leftarrow \texttt{LOCKED}$ with a $\Delta_{lock}$\;

     send to $S:\langle \textsf{EXEC}, cid, rid, req, \delta \rangle$\;
    $k \leftarrow k+1$\;
}

\vspace{0.8em}
\on{$S.\mathtt{execRequest}(cid, rid, req, \delta)$}{
    $res \leftarrow \mathtt{Execute}(req)$\;
    $t \xleftarrow{\$} \mathbb{Z}_q$\;
    $T \leftarrow t\cdot G$\;

    $h \leftarrow \Hash(res)$\;
    $m \leftarrow \Hash(cid \parallel rid \parallel \delta \parallel h)$\;

    $\hat{\sigma}_S \leftarrow \mathtt{pSign}(sk_S, m, T)$\;
    $EncRes \leftarrow \mathtt{Enc}_t(res)$\;

    send to $\mathcal{U}:\langle \textsf{REPLY}, h, T, \hat{\sigma}_S, EncRes\rangle$\;
}

\columnbreak
\setcounter{AlgoLine}{23}

\on{$\mathcal{U}.\mathtt{onExecReply}(h, T,\hat{\sigma}_S, EncRes)$}{
    $(\delta,req,\_,\_,\_) \leftarrow \mathtt{Pend}[cid,rid]$\;
    $m \leftarrow \Hash(cid \parallel rid \parallel \delta \parallel h)$\;
    \assert $\mathtt{pVerify}(pk_S, m, T, \hat{\sigma}_S)$\;

    $\mathtt{Pend}[cid,rid] \leftarrow (\delta,req,EncRes,\hat{\sigma}_S,T)$\;

    $\varsigma \leftarrow \texttt{PENDING}$ in $\Gamma_{cid}$\;
    $\sigma_U \leftarrow \mathtt{Sign}_U(m, \delta)$\;

     send to $S:\langle \textsf{AUTH}, \sigma_U\rangle$\;
}

\vspace{0.8em}
\on{$S.\mathtt{revealSecret}(\sigma_U)$}{
    $m \leftarrow \Hash(cid \parallel rid \parallel \delta \parallel h)$\;
    \assert $\mathtt{Verify}(pk_u,m,\delta,\sigma_U)$ \;
    \tcp*[l]{Preferred off-chain reveal}
    $send\ to\ \mathcal{U}:\langle \textsf{REVEAL}, t\rangle$\;
    $\mathtt{Wait}(\Delta_{\mathsf{ack}})$\;

    \eIf{$\Delta_{\mathsf{ack}}$ received}{
        \Return \texttt{OFFCHAIN\_REVEAL}\;
    }
    {\tcp*[l]{Fallback: on-chain settlement}
    $\sigma_S \leftarrow \mathtt{AdaptSig}(\hat{\sigma}_S, t)$\;
    $\mathtt{forceCloseASC}(cid, \sigma_S)$\;
    $t \gets \mathcal{U}.\mathtt{extract}(\hat{\sigma}_S, \sigma_S, T)$\;
    \Return \texttt{ONCHAIN\_REVEAL}\;}
}

\columnbreak
\setcounter{AlgoLine}{45}

\on{$\mathcal{U}.\mathtt{onSecretReveal}(t)$}{
    $(\delta,req,EncRes,\hat{\sigma}_S,T) \leftarrow \mathtt{Pend}[cid,rid]$\;
    $res \leftarrow \mathtt{Dec}_t(EncRes)$\;

    $(C,S,\mathbf{B},\varsigma,k) \leftarrow \Gamma_{cid}$\;
    $b^{C}_{locked} \mathrel{-}= \delta ;b^{S}_{free} \mathrel{+}= \delta$\;
    remove $\mathtt{Pend}[cid,rid]$\;
    $\varsigma \leftarrow \texttt{OPEN}$ in $\Gamma_{cid}$\;

    issue receipt $\Delta_{\mathsf{ack}}$ to $S$\;
    deliver $res$ to $C$\;
    
}

\end{multicols}
\end{algorithm*}

\subsection{Atomic Exchange Protocol}
\label{sec:design_pay_deliver}
With an established Atomic Service Channel, A402
enables high-frequency off-chain service interactions. The remaining challenge is to enforce the atomicity between service execution, payment, and result delivery for each service request. We formalize this requirement by defining the notion of \emph{atomic exchange}.


\begin{definition}[Atomic Exchange]
An atomic exchange between a vault $\mathcal{U}$ and a service provider
$S$ satisfies the following property: for any execution of the protocol involving an honest party and a malicious counterparty, the
protocol terminates in exactly one of the following states:

\begin{itemize}
    \item \emph{Success:} $\mathcal{U}$ obtains the service result
    $res$, and $S$ obtains the payment $\delta$;
    \item \emph{Abort:} $\mathcal{U}$ retains the payment $\delta$, and
    $S$ does not obtain $\delta$ (and may retain $res$ or no execution
    occurs).
\end{itemize}

Formally, for any adversary controlling one party,
the probability that the protocol terminates in the following \emph{forbidden states}
is negligible:
\[
\begin{aligned}
\Pr\big[\, 
    &(S\ \text{obtains}\ \delta \wedge \mathcal{U}\ \text{does not obtain}\ res) \\
    \lor\; 
    &(\mathcal{U}\ \text{obtains}\ res \wedge S\ \text{does not obtain}\ \delta)
\,\big]
\le \mathtt{negl}(\lambda).
\end{aligned}
\]
\end{definition}



    

\begin{figure}[h]
    \centering
    \includegraphics[width=\linewidth]{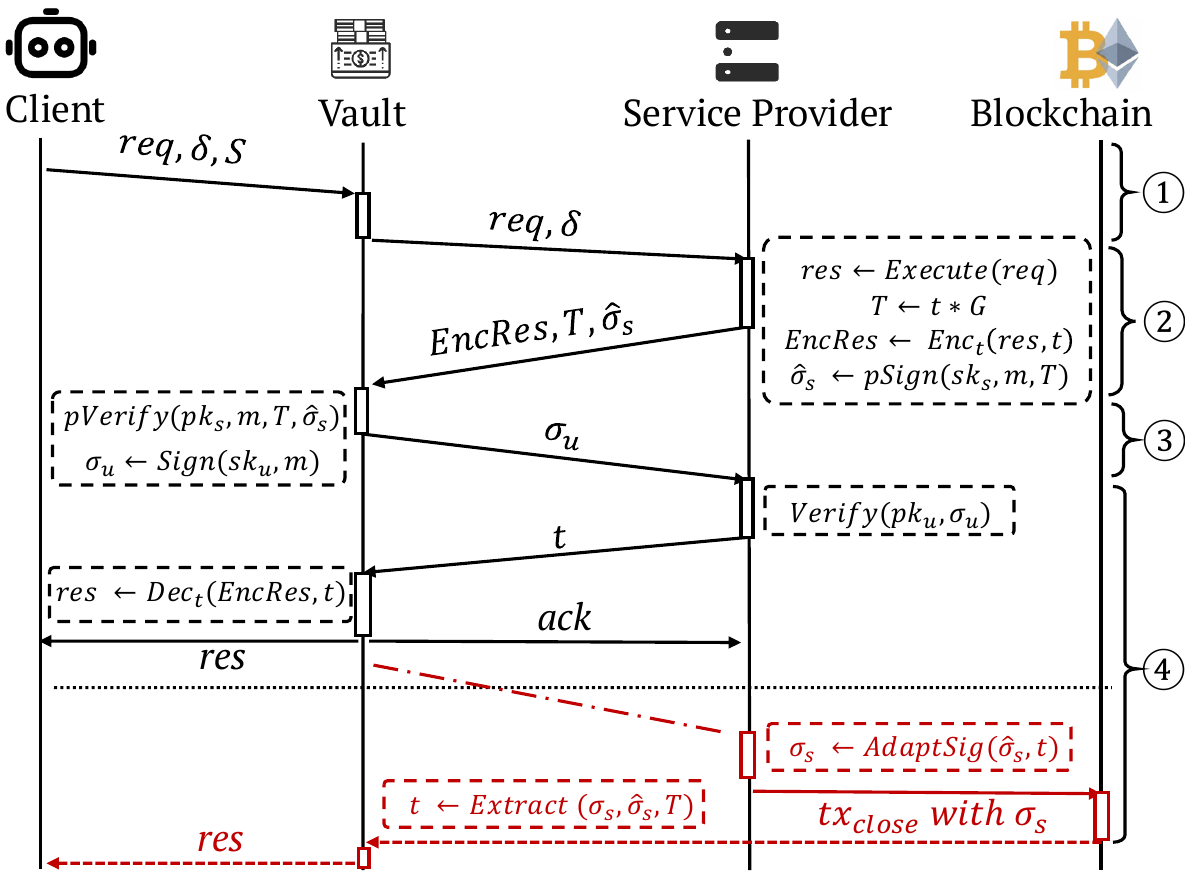}
    \caption{The workflow of atomic exchange}
    \label{fig:workflow}
\end{figure}

\myparagraph{Atomic Exchange using TEE-assisted Adaptor Signatures}
A402 realizes atomic exchange using \emph{TEE-assisted adaptor signatures}, as
illustrated in Figure~\ref{fig:workflow} and formalized in
Algorithm~\ref{alg:atomicexchange}.
The protocol proceeds in four phases and enforces
\emph{Exec-Pay-Deliver atomicity} for each service interaction.




\noindent\emph{\circled{1} Request Submission and Asset Locking.}
As shown in Figure~\ref{fig:workflow}~\circled{1}, the exchange is initiated
when the vault $\mathcal{U}$ submits a service request $req$ on behalf of the
client $C$. 
The $\mathcal{U}$ first verifies that the ASC is in the \texttt{OPEN} state and
optimistically escrows the payment amount $\delta$ by moving it from
$b^{C}_{free}$ to $b^{C}_{locked}$
(Algorithm~\ref{alg:atomicexchange}, lines~4-8).
A request identifier $rid=(cid,k)$ is generated (line 7) and
the ASC then transitions to the \texttt{LOCKED} state, ensuring that no concurrent requests are accepted while an execution is in flight.
To preserve liveness and prevent the ASC from being indefinitely stalled by a non-responsive $S$,
the $\mathcal{U}$ enforces a timeout $\Delta_{lock}$. If $\mathcal{U}$ fails to receive a valid $\textsf{REPLY}$ from $S$ within $\Delta_{lock}$, the ASC automatically reverts to the \texttt{OPEN} state, effectively expiring the $cid$ and restoring ASC availability (line 10).
After that, the tuple
$(cid,rid,req,\delta)$ is forwarded to the $S$ via a
secure channel (line 11).



\noindent\emph{\circled{2} TEE-assisted Execution and Adaptor-Signature Payment Commitment.}
Upon receiving the request, the $S$ executes $req$ inside its
attested TEE, obtains the result
$res$ (line 15) and then commits payment via an adaptor signature (lines 16-20), as illustrated in
Figure~\ref{fig:workflow}~\circled{2}.
Because both the execution logic and the payment commitment logic are enforced inside a TEE, payment commitment is generated if and only if the $req$ has been correctly executed.
Moreover, to bind execution to payment commitment, $S$ samples a  secret witness $t \in \mathbb{Z}_q$ inside the TEE and derives the corresponding public statement $T = t \cdot G$ (lines~16-17).
The execution result is encrypted using $t$ to produce $EncRes$, and an adaptor pre-signature
$\hat{\sigma}_S \leftarrow \mathtt{pSign}(sk_S, m, T)$
is generated over the payment message $m$ (lines~19-21). The $S$ then returns an \emph{execution-bound payment commitment}
$\Pi = (EncRes, T, \hat{\sigma}_S)$ to $\mathcal{U}$ (line 22).

Crucially, the TEE enforces \emph{execution-payment commitment atomicity}. The witness $t$ is generated and sealed inside the TEE and cryptographically
binds the encrypted execution result $EncRes$ to the payment commitment $\Pi$. Thus, a payment commitment cannot be generated without performing the service execution.


\noindent\emph{\circled{3} Execution Verification and Conditional Payment.}
Upon receiving $\Pi$, the $\mathcal{U}$ reconstructs the payment message
$m$ and verifies the adaptor pre-signature $\hat{\sigma}_S$ via $\mathtt{pVerify}$
(lines~26-28).
A successful verification confirms that the $S$ has produced a
valid $\Pi$ under the statement $T$.
$\mathcal{U}$ then transitions the ASC to
\texttt{PENDING}, and issues a conditional payment signature
$\sigma_U$ to the $S$ (lines~29--31).
At this point, payment is approved but cannot be finalized without revealing the corresponding witness $t$.

\noindent\emph{\circled{4} Payment Finalization and Result Delivery.}
After validating $\sigma_U$, $S$ reveals the $t$ and finalizes this exchange
through one of two paths (lines 34-35), as illustrated in
Figure~\ref{fig:workflow}~\circled{4}.
In the off-chain path, if $\mathcal{U}$ is responsive and cooperates, $S$ reveals the witness $t$
directly to $\mathcal{U}$ over the secure channel
(lines~36--40).
Upon receiving $t$,
the $\mathcal{U}$ decrypts $EncRes$, transfers $\delta$ from
$b^{C}_{locked}$ to $b^{S}_{free}$, updates the channel state
to \texttt{OPEN}, and delivers the plaintext result $res$ to the $C$ (lines 47-54), thereby finalizing the payment and enabling result delivery.
If the $\mathcal{U}$ fails to respond within time $\Delta_{ack}$ or the $S$ chooses to withhold the witness $t$ off-chain, 
$\mathcal{S}$ completes the adaptor signature
$\sigma_S \leftarrow \mathtt{AdaptSig}(\hat{\sigma}_S,t)$ (line 41) and invokes
$\mathtt{forceCloseASC}(\sigma_S)$ to trigger on-chain settlement (lines 42-43).
Since on-chain settlement transaction reveals the completed adaptor signature $\sigma_S$ (lines 61 of Algorithm~\ref{alg:asc}), $\mathcal{U}$ can therefore extract $t$ and decrypt the execution result $res$ (line~43), ensuring that payment finalization remains inseparable from result delivery even under adversarial conditions.




Note that adaptor signatures enforce \emph{payment finalization - result delivery atomicity}.
Any attempt by $S$ to finalize payment—either off-chain or on-chain—necessarily reveals the witness $t$, which enables $\mathcal{U}$ to recover the result $res$.


\myparagraph{Exec-Pay-Deliver Atomicity}
Using TEE-assisted adaptor signature, A402 
enforces Exec-Pay-Deliver Atomicity.
We decompose Exec-Pay-Deliver atomicity into two properties:
\emph{Exec-Pay atomicity} and \emph{Pay-Deliver atomicity}. TEEs bind execution to payment commitment, and adaptor signatures bind payment finalization to result delivery.

\begin{algorithm*}[t]
\caption{Privacy-Preserving Liquidity Vault}
\label{alg:a402-liquidity}
\scriptsize
\begin{multicols}{3}
\noindent\textbf{Vault Account State:}\;
\quad $\mathcal{V}_p = (v_p^{free}, v_p^{locked})$ for $p\in\{C,S\}$\;


\vspace{0.6em}

\on{$p.\mathtt{initVault}(\pi)$}{
    \assert $\mathcal{U} \gets \mathtt{initCommittee}(\pi)$\;
    issue $tx_{init}$ from $p$\;
    \assert $\mathcal{L}.\mathtt{exec}(tx_{init})$\;
    $v_p^{free} \mathrel{+}= tx_{init}.val$\;
}

\vspace{0.8em}

\on{$C.\mathtt{reqOpenVaultASC}(S,\mathcal{U}, v)$}{
    \assert $\neg \mathcal{U}.\mathtt{hasASC}(C,S)$\;
    \If{$\neg\mathcal{U}.\mathtt{hasSC}(S, SC)$}{
        $\mathcal{U}.\mathtt{establishSC}(S,SC)$\;
    }
    $v_C^{free} \mathrel{-}= v$\;
    $(cid, \Gamma_{cid}) \gets \mathtt{initVaultASC}(C, S, v)$\;
    \Return $\Gamma_{cid}$\;
}

\vspace{0.8em}

\columnbreak
\setcounter{AlgoLine}{17}

\on{$p.\mathtt{reqCloseVaultASC}(cid)$}{
    $\Gamma \gets \mathtt{ASC}[cid]$\;

    \If{$\Gamma.\varsigma = \texttt{OPEN}$}{
        $\mathcal{U}.\mathtt{closeVaultASC}(\Gamma)$
    }
    \If{$\Gamma.\varsigma \in \{\texttt{LOCKED},\texttt{PENDING}\}$}{
        $\mathcal{U}.\mathtt{WaitOrObserveOnChain}(cid)$\;
    }
}

\vspace{0.8em}

\on{$p.\mathtt{reqSettleVault}()$}{
    \assert $v_p^{free} > 0$\;
    $tx_{settle} \gets \mathcal{U}.\mathtt{buildSettleTX}(p, v_p^{free})$\;
    \assert $\mathcal{L}.\mathtt{exec}(tx_{settle})$\;
    $v_p^{free} \leftarrow 0$\;
}

\columnbreak
\setcounter{AlgoLine}{33}
\vspace{0.8em}

\on{$p.\mathtt{forceSettleVault}(\mathcal{V}_p)$}{
    $tx_{init} \gets \mathtt{initForceExitTx}(\mathcal{V}_p)$\;
    \assert $\mathcal{L}.\mathtt{exec}(tx_{init})$\;
    $\mathcal{L}.\mathtt{StartDisputeTimer}(\Delta_{settle})$\;

    \If{$\Delta$ expired and no valid challenge}{
        $tx_{final} \gets \mathtt{finalizeForcExitTx}(tx_{init})$\;
        \assert $\mathcal{L}.\mathtt{exec}(tx_{final})$\;
        $v^{free}_p \gets 0$
    }
}

\end{multicols}
\end{algorithm*}

\subsection{Privacy-Preserving Liquidity Vault}
\label{sec:design_privacy}
While Atomic Service Channels significantly reduce on-chain interaction during steady-state operation, their creation and closure still require on-chain
transactions, which introduce non-trivial privacy leakage.
In particular, the ASC creation and settlement transactions
$tx_{create}$ and $tx_{close}$ publicly reveal the interaction relation between the client $C$ and the service provider $S$, as well as the exact channel capacity.
To mitigate this leakage, A402 introduces a \emph{Privacy-Preserving Liquidity Vault}, a TEE-backed protocol that decouples individual ASC lifecycles from 
per-ASC on-chain transactions.
Specifically, assets are aggregated into an off-chain liquidity pool managed by an attested vault. Algorithm~\ref{alg:a402-liquidity} formalizes the workflow of liquidity vault, in which channel creation, execution, and closure are handled off-chain.




\myparagraph{Vault Account Model}
A402 maintains a vault-based account model to track participant liquidity.
For each participant $p \in \{C,S\}$, the liquidity vault maintains a private
account state
\[
\mathcal{V}_p = (v_p^{free}, v_p^{locked}),
\]
which is stored and updated entirely inside the vault’s attested TEE.
The free balance $v_p^{free}$ represents assets available for
funding new ASCs or for batched on-chain settlement, while $v_p^{locked}$ captures liquidity temporarily reserved by active
ASCs.
All balance transitions are enforced by verified TEE code and are never exposed on-chain, ensuring that per-channel balances and counterparty relationships remain hidden from external observers. 

\myparagraph{Liquidity Vault Initialization}
The liquidity vault is initialized via the
$\mathtt{initVault}$ interface (Algorithm~\ref{alg:a402-liquidity}, lines~3-7), which sets up a TEE-backed committee and transfers user assets into a committee-controlled address.
Upon invocation, the user specifies a policy parameter $\pi$ defining a
$m$-out-of-$n$ committee rule (line 3).
The vault executes $\mathtt{initCommittee}(\pi)$ to deterministically instantiate
a committee of $n$ attested TEE instances and configure the corresponding threshold governance logic inside the TEE (line 4).
After committee initialization, $p$ submits an on-chain deposit transaction
$tx_{init}$ to a $\mathcal{U}$-controlled address.
Once the transaction is confirmed, $\mathcal{U}$ credits the deposited assets to the free balance $v_p^{free}$ (lines 6-7).
The assets are then decoupled from any specific ASC on-chain and can be flexibly allocated to multiple ASCs off-chain under
$\mathcal{U}$’s control.

\myparagraph{Private ASC Creation}
To create an ASC under the liquidity-vault model, the client $C$ invokes $\mathtt{reqOpenVaultASC}$
(lines~9-17).
The $\mathcal{U}$ first ensures that no active ASC exists between $C$ and
$S$, and establishes a secure channel with $S$ if needed (lines 10-12).
The requested channel capacity $v$ is debited from $C$’s free vault balance
without generating any on-chain transaction (line 14).
The vault then generates a fresh ASC identifier $cid$ and initializes the
ASC state $\Gamma_{cid}$ with status \texttt{OPEN} (line 15).
As a result, ASC creation becomes a purely off-chain operation, leaving no public on-chain transaction that links $C$ and $S$.


\paragraph{Private ASC Closure.}
Either party may request a cooperative ASC closure via
$\mathtt{reqCloseVaultASC}$
(lines~18-26).
If the ASC is in the \texttt{OPEN} state, $\mathcal{U}$ invokes $\mathtt{closeVaultASC}$, releases the ASC’s free
assets and credits them back to the corresponding vault accounts before marking
the ASC as \texttt{CLOSED} (lines 20-21).
If the ASC is in the \texttt{LOCKED} or \texttt{PENDING} state, the vault
defers closure and waits for the in-flight request to complete or for an
on-chain force-close transaction to be observed (line 24).
Importantly, assets locked in an active ASC remain managed by the standard ASC protocol (Algorithms~\ref{alg:asc} and \ref{alg:atomicexchange}) and can always be reclaimed via on-chain force-close transactions.
As long as the vault $\mathcal{U}$ remains live, both $C$ and $S$ are economically disincentivized from invoking on-chain force-close, since cooperative off-chain closure is strictly more efficient and privacy-preserving.
As a result, privacy-preserving settlement does not compromise the \emph{Trust-Minimized Asset Security}.

\paragraph{Liquidity Vault Settlement.}
When a party $p$ (\ie, $C$ or $S$) intends to withdraw assets from $\mathcal{U}$, it
invokes $\mathtt{reqSettleVault}()$
(lines~27-32).
The $\mathcal{U}$ aggregates the participant $p$'s free balance ($v^{free}_{p}$), including assets released from previously closed ASCs, and constructs a single on-chain transaction $tx_{settle}$ that transfers
the $v_p^{free}$ back on chain (lines 29-30).
By consolidating multiple off-chain ASC settlements into a single on-chain transaction, $tx_{settle}$ decouples individual ASCs from on-chain asset transfers, which reduces the traceability of ASC settlements.



\paragraph{Force Liquidity Vault Settlement.}
To tolerate $\mathcal{U}$ unavailability, \textsc{A402} provides an on-chain escape mechanism via
$\mathtt{forceSettleVault}$
(Algorithm~\ref{alg:a402-liquidity}, lines~34-43).
The participant $p$ submits a force-settle initialization transaction $tx_{init}$,
which opens an on-chain dispute window $\Delta_{settle}$ (lines 35-37).
If $\mathcal{U}$ fails to submit a conflicting state during this window, $p$
finalizes the withdrawal after timeout (lines 38-40)\footnote{Similarly, $\mathcal{U}$ can utilize a corresponding \emph{watchtower} to monitor and react to force-settle transactions during the challenge period.}.
This mechanism guarantees asset security even under a malicious or offline $\mathcal{U}$,
while preserving the privacy benefits of aggregating off-chain ASC settlements.

\section{Security Analysis}
\label{sec:security}


\myparagraph{Trust-Minimized Asset Security}
A402 guarantees asset security without relying on the availability or honest
behavior of either the vault $\mathcal{U}$ or the service provider $S$.
In the standard mode, ASC assets are locked and settled by the underlying blockchain. If $\mathcal{U}$ becomes unresponsive or censors off-chain interactions, both $C$ and $S$ can unilaterally initiate an on-chain closure. Under the liquidity vault mode, assets are managed off-chain within $\mathcal{U}$ but secured by verifiable TEE-enforced state transitions. To prevent permanent asset locking, any participant $p \in \{C, S\}$ can unilaterally trigger an on-chain force-settle if $\mathcal{U}$ becomes unresponsive or adversarial.

\myparagraph{Exec-Pay-Deliver Atomicity}
A402 ensures atomicity between service execution, payment, and result delivery, protecting against malicious clients, service providers, and vaults. A402 uses TEE-assisted adaptor signatures to enforce atomicity. The TEE ensures that the payment commitment is generated only after service execution, preventing manipulation by $S_{adv}$ before service execution. The adaptor signature binds payment finalization to the revelation of witness $t$ embedded in the encrypted result, ensuring that $S_{adv}$ cannot finalize payment without delivering the result. Against malicious clients ($C_{adv}$), A402 prevents free-riding by enforcing payment authorization and state transitions within the vault’s TEE, blocking double-spending, payment blocking, or unauthorized result access. Against malicious vaults ($\mathcal{U}_{adv}$), TEE ensures that payment cannot be finalized without a valid adaptor signature from the $S$. If the vault becomes unresponsive, the $S$ and $C$ can unilaterally enforce settlement on-chain.

\myparagraph{Unlinkability}
In the privacy-preserving liquidity vault mode, A402 protects against an adversary that observes the blockchain.
Because ASC creation, updates, and closure occur entirely off-chain, on-chain transactions expose only vault initialization and settlement.
An observer cannot link a specific on-chain transaction to a particular ASC creation or closure, nor infer the
frequency, volume, or interactions between clients and service providers.


\section{Implementation}
\label{sec:imple}

\subsection{Off-chain Component}

The off-chain components of A402 consist of three entities: the \emph{vault ($\mathcal{U}$)}, the \emph{service provider ($S$)}, and the \emph{client ($C$)}. Both the $\mathcal{U}$ and the $S$ are implemented in C++17 and deployed inside AMD SEV-SNP protected virtual machines, forming the TEE instance. The $C$ is considered an untrusted party.

\myparagraph{Vault} We implemented $\mathcal{U}$ as a scalable component that can be deployed either as a single TEE or as a replicated cluster of multiple TEEs. In the clustered configuration, our implementation uses the Raft consensus protocol\cite{ongaro2014search} with a leader-follower architecture.

\myparagraph{Service Provider} To decouple protocol overhead from specific workloads, we simulate service execution with a fixed 200 ms delay in $S$. The adaptor signature module is implemented inside the TEE using a \emph{Schnorr adaptor signature scheme}\cite{gerhart2024foundations}.


\subsection{On-chain Component}
\label{subsec:onchain}
We implement the on-chain support for ASCs on a locally controlled Ethereum test network and Bitcoin test network (Appendix~\ref{appendix:bitcoin}). 

\myparagraph{Ethereum: ASC Manager Contract}
We implement the on-chain component of A402 as an \textbf{ASC Manager} smart contract deployed on an Ethereum test network using the Hardhat toolchain (Hardhat~v2.19.0, ethers.js~v6.9.0, and hardhat-toolbox~v4.0.0).
The contract manages the full lifecycle of ASCs. Channel creation and cooperative closure are provided via \texttt{createASC} and \texttt{closeASC}. To handle adversarial scenarios, the contract exposes unilateral client closure via \texttt{initForceClose}, \texttt{finalForceClose} and unilateral service provider closure via \texttt{forceClose}.
In liquidity-vault mode, the contract supports vault initialization and settlement via \texttt{initVault} and \texttt{settleVault}, together with force-settlement interfaces (\texttt{initForceSettle}, \texttt{finalForceSettle}) that guarantees asset recoverability in the presence of vault failures.
A practical challenge is that Ethereum does not provide native \texttt{secp256k1} Schnorr verification precompiles.
To enable on-chain verification of adaptor signatures, we integrate an optimized solidity schnorr verification library\footnote{\url{https://github.com/noot/schnorr-verify}}. A detailed mapping between these on-chain interfaces and the ASC protocol interfaces is provided in Table~\ref{tab:api_summary} in the Appendix~\ref{appendix:interface}.


\myparagraph{Bitcoin} Detailed implementation of the Bitcoin on-chain component can be found in the Appendix~\ref{appendix:bitcoin}.

\section{Evaluation}
\label{sec:eval}
\subsection{Setup}



All experiments are conducted on a server equipped with an AMD EPYC 9965 CPU (192 cores) and 512GB RAM. The machine hosts all off-chain components of A402.
To isolate resource contention, CPU cores are statically partitioned across system components. The service provider $S$ is allocated 100 CPU cores, while the vault $\mathcal{U}$ is allocated 64 CPU cores. In the clustered configuration, each replica is assigned 8 CPU cores.
We emulate network communication between $C$, $S$, and $\mathcal{U}$ by introducing a fixed 10 ms network delay per link. This delay is held constant across all experiments.
To decouple protocol overhead from application-specific workloads, the service provider introduces a fixed 200 ms processing delay per request. The service provider $S$ maintains 100 ASCs with $\mathcal{U}$, each pre-funded with sufficient client assets to avoid liquidity bottlenecks during experiments. We implement a client workload generator that simulates 100 concurrent clients. Each client continuously issues service requests to $S$ through $\mathcal{U}$. The total number of requests ranges from 100 to 10,000 and is evenly distributed across the 100 pre-established ASCs.

\subsection{On-chain Cost}
We evaluate the on-chain transaction cost of A402 on both Bitcoin and Ethereum and compare it with x402. In x402, each service request requires an on-chain payment transaction, resulting in \emph{linear on-chain cost} with respect to the number of requests ($O(n)$). In contrast, A402 amortizes on-chain interactions across multiple requests.
In standard mode, A402 requires only channel creation and closure transactions. In liquidity vault mode,  A402 requires vault initialization and settlement transactions. As a result, the number of on-chain transactions in A402 remains constant with respect to the number of off-chain service requests($O(1)$).

\begin{table}[t]
\centering
\small
\caption{\bf Ethereum Gas Consumption: Detailed Interface Costs. \normalfont{Gas Price: 20 Gwei, ETH Price: \$2,000.}}
\label{tab:eth_detailed_cost}
\resizebox{\linewidth}{!}{
\begin{tabular}{lrr}
\toprule
\textbf{Interface / Operation} & \textbf{Gas Used} & \textbf{Cost (USD)} \\
\midrule
\multicolumn{3}{l}{\textbf{\textit{A402: Standard Channel Mode (ASC)}}} \\
$\mathcal{U}.\texttt{createASC}$ & 26,050 & \$1.04 \\
$\mathcal{U}.\texttt{closeASC}$ & 57,637 & \$2.31 \\
\cmidrule(lr){1-3}
\textit{Total Cost ($n$ request)} & \textit{\textbf{83,687}} & \textit{\textbf{\$3.35}} \\
\cmidrule(lr){1-3}
\multicolumn{3}{l}{\textit{\footnotesize --- Force Close Operations ---}} \\
$S.\texttt{forceClose}$ & 116,650 & \$4.67 \\
$C.\texttt{initForceClose}$ & 51,980 & \$2.08 \\
$C.\texttt{finalForceClose}$ & 62,787 & \$2.51 \\
\midrule
\multicolumn{3}{l}{\textbf{\textit{A402: Liquidity Vault Mode}}} \\
$\mathcal{U}.\texttt{initVault}$ & 23,846 & \$0.95 \\
$\mathcal{U}.\texttt{settleVault}$ & 27,490 & \$1.10 \\
\cmidrule(lr){1-3}
\textit{Total Cost ($n$ request)} & \textit{\textbf{51,336}} & \textit{\textbf{\$2.05}} \\
\cmidrule(lr){1-3}
\multicolumn{3}{l}{\textit{\footnotesize --- Force Settle Operations ---}} \\
$C/S.\texttt{initForceSettle}$ & 114,237 & \$4.57 \\
$C/S.\texttt{finalForceSettle}$ & 23,502 & \$0.94 \\
\midrule
\multicolumn{3}{l}{\textbf{\textit{Baseline: x402}}} \\
Approve (One-time) & 22,580 & \$0.90 \\
TransferFrom (Single Request) & 23,530 & \$0.94 \\
\cmidrule(lr){1-3}
\textit{Total Cost ($n$ request)} & \textit{\textbf{22,580 + 23,530 * n}}\textsuperscript{\dag} & \textit{\textbf{\$0.9 + \$0.94 * n}} \\

100 Requests & 2,375,580 & \$95.02 \\
\bottomrule
\multicolumn{3}{l}{\footnotesize \textsuperscript{\dag} Calculated as: $\text{Gas}_{\text{Approve}} + (n \times \text{Gas}_{\text{TransferFrom}})$.}
\end{tabular}}
\end{table}

\myparagraph{Transaction Cost on Ethereum}
We measure the gas consumption of A402 using the contract interfaces described in Section~\ref{subsec:onchain}.
Both A402 and x402 use ERC20 tokens as the payment asset to ensure a fair comparison. 
Table~\ref{tab:eth_detailed_cost} summarizes the measured costs.

In standard mode, ASC creation and cooperative closure cost 26,050 and 57,637 gas, resulting in a total lifecycle cost of 83,687 gas (\$3.35) per channel.
In liquidity vault mode, vault initialization and settlement cost 51,336 gas in total (\$2.05). These costs are incurred once per channel (or vault) lifecycle and are therefore amortized across all off-chain service requests.
To guarantee asset recovery under adversarial conditions, A402 provides force-closure paths. The service provider can enforce closure directly via $S.\texttt{forceCloseASC}$ (116,650 gas), while client-driven force-closure via $C.\texttt{initForceClose}$ and $C.\texttt{finalForceClose}$ costs 23,846 and 27,490 gas. In liquidity vault mode, force settlement via \texttt{initForceSettle} and \texttt{finalForceSettle} costs approximately \$5.51. The above interfaces serve strictly as a worst-case fallback for asset recovery and are expected to be invoked with negligible frequency. Even when triggered, their costs remain bounded and occur at most once per channel/vault lifecycle.

\myparagraph{Comparison with x402} We model x402 using a \emph{single ERC20 approval} transaction followed by repeated \texttt{transferFrom} calls, representing the best-case deployment of x402. 
Specifically, x402 requires an on-chain transfer for every request. Its total gas cost is: $22,580 + 23,530*n$, where the 22,580 is the gas cost of the one-time approval transaction. As a concrete example, processing \emph{100 requests} costs approximately \$95.02 in x402, while A402 incurs only the one-time lifecycle cost of 
\$3.35 (standard mode) or \$2.05 (vault mode), representing a cost reduction of approximately \emph{28$\times$–46$\times$}.

\myparagraph{Transaction Cost on Bitcoin} Detailed results can be found in the Appendix~\ref{appendix:bitcoin}.

\subsection{Performance Evaluation}
\label{subsec:performance}
We evaluate the off-chain performance of A402’s ASC, focusing on \textbf{throughput} and \textbf{end-to-end latency}. 
Throughput is measured as completed requests per second (RPS), and latency is measured as the end-to-end request completion time.
Our evaluation consists of two workloads: (i) varying service request volumes with a single vault, and (ii) scaling the number of vault replicas under a fixed request volume (details in the Appendix~\ref{appendix:replica}).

\begin{figure}
    \centering
    \includegraphics[width=0.85\linewidth]{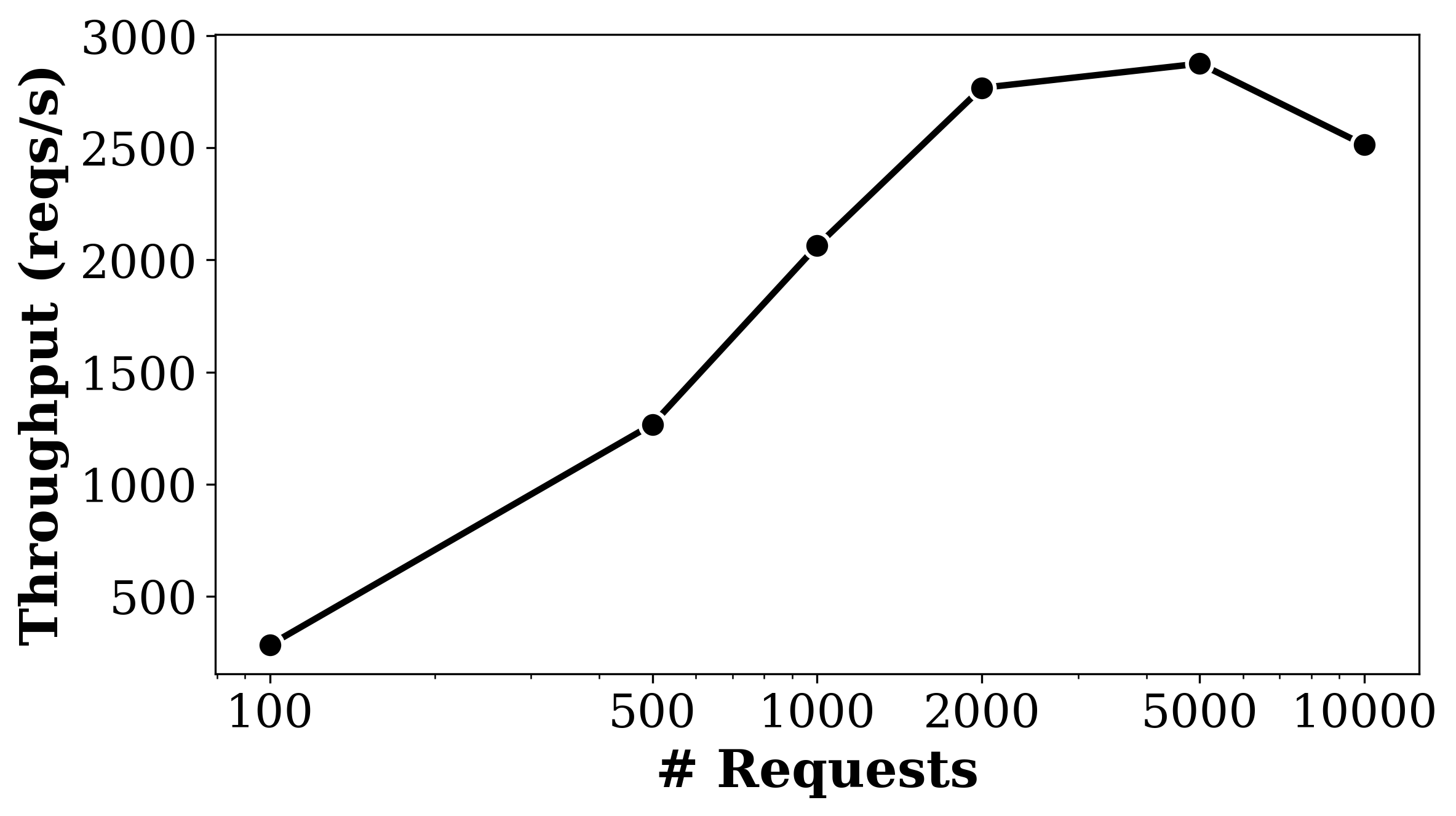}
    \caption{The throughput of A402 with varying requests}
    \label{fig:throughput1}
\end{figure}

\begin{figure}
    \centering
    \includegraphics[width=0.85\linewidth]{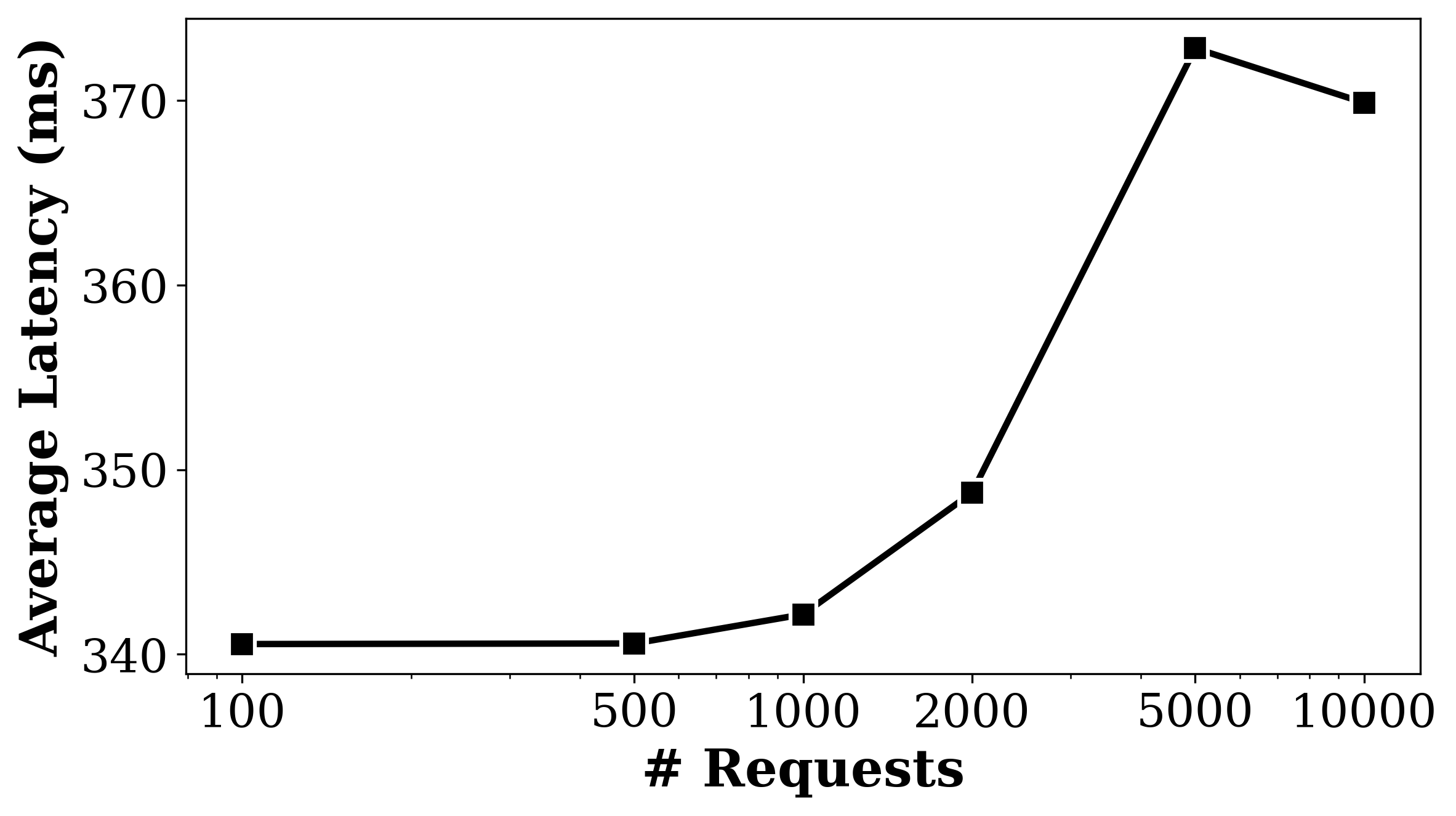}
    \caption{The latency of A402 with varying requests.}
    \label{fig:Latency1}
\end{figure}

\myparagraph{Scalability with Request Volume}
Figures~\ref{fig:throughput1} and \ref{fig:Latency1} show the throughput and latency of ASC, respectively, as the number of concurrent requests increases from 100 to 10,000. 
Throughput scales significantly and reaches a peak of 2,875 RPS at 5,000 concurrent requests.
We observe a mild saturation phase from 5,000 to 10,000 requests due to resource contention. As shown in Figure~\ref{fig:Latency1}, average latency remains stable, increasing only by 8.6\%, from 340.56 ms to 369.86 ms, as the request volume scales. This latency is largely determined by the simulated service execution time (200~ms) and network delay (10 ms), indicating that the additional overhead introduced by ASC is small. These results indicate that ASC effectively handles high request volumes while maintaining low latency.


\paragraph{Comparison with x402.}
For fairness, we compare against an idealized deployment of x402 and consider only the  throughput and latency of on-chain payment, excluding additional real-world overheads such as network delay, request execution.
Because x402 ultimately relies on on-chain payment, its performance is bounded by underlying blockchain.
Our prototype achieves a peak throughput of 2,875 RPS, exceeding the payment throughput of Ethereum ($\sim$30 TPS\cite{chainspect_ethereum}) and Bitcoin ($\sim$7 TPS\cite{chainspect_bitcoin}) by 95$\times$ and 410$\times$, respectively. In terms of latency, blockchain confirmation typically requires seconds to minutes ($\sim$12.8 s on Solana\cite{chainspect_solana}, $\sim$13 min on Ethereum\cite{chainspect_ethereum}, and $\sim$ 60 min on Bitcoin\cite{chainspect_bitcoin}), whereas A402 completes the full service request in 0.34–0.37 s on average—over 40$\times$ faster than Solana and several orders of magnitude faster than Ethereum and Bitcoin.

\myparagraph{Horizontal Scalability} Unlike x402, whose performance is strictly bound by the underlying blockchain, A402 features a scale-out architecture where capacity grows linearly with additional hardware resources.
Notably, a single deployment of our prototype already achieves a throughput of 2,875 RPS, comparable to high-performance blockchains like Solana (1,600-3,500 TPS\cite{chainspect_solana}).
This implies that by simply provisioning additional vaults and service providers, A402 can support high-frequency agentic interactions at a scale far exceeding that of x402.

\section{Applications}
\label{sec:appli}
This section provides a brief overview of several new and novel applications enabled by A402.



\myparagraph{Request-Centric Micropayments in Web 2.0 Service} 
$\textsc{A402}$ shifts Web 2.0 services from an identity-centric mode (\eg, persistent API keys and subscriptions) to a request-centric payment. 
Specifically, $\textsc{A402}$ treats each service invocation as an economic event: a client no longer pays for general "platform access" but rather for the service request, \ie, the execution of a function $f(x) \to y$.

\emph{\textbf{Pre-A402}: "I pay for access to the platform, hoping the platform will
execute my request correctly.".}

\emph{\textbf{Post-A402}: "I pay only for this attested execution of $f(x)$ that
produces result $y$". }

For service providers, this shift significantly lowers the operational barrier to entry. By removing the need for user database maintenance, the service provider needs only to expose an attested TEE endpoint.


\myparagraph{Autonomous agentic Commerce via Pay-for-Verified-Execution} 
$\textsc{A402}$ transforms Web 2.0 services into a trustless marketplace where verifiable execution is the fundamental unit of exchange. By leveraging ASCs, autonomous AI agents can programmatically negotiate and consume computational resources—such as real-time inference or private data queries—without the risk of service non-delivery. 
This $\textsc{A402}$ ensures that payment is finalized only upon the delivery of a valid result, enabling agents to securely compose complex, cross-service provider workflows providing the essential financial rail for a scalable, self-directed agentic commerce.



\myparagraph{High-Frequency Trustless Metering for Edge and IoT}
Edge and IoT environments require millisecond-level latency and high-concurrency interactions between mutually untrusted machines. x402 fails to support such payments due to blockchain limitations, while subscription models are too coarse-grained for sporadic machine workloads. In contrast, A402 enables trustless off-chain metering and settlement with millisecond-level overhead, making it well suited for latency-sensitive scenarios such as 5G routing and CDN caching.



\section{Related Work}
\label{sec:rw}

\myparagraph{Payment Channel}
Prior research on payment channel has largely focused on enhancing privacy\cite{heilman2017tumblebit,green2017bolt,qin2023blindhub} and capital efficiency\cite{khalil2017revive,ge2024shaduf,wu2022cycle}. 
%
Most closely related to our work are Teechan\cite{lind2017teechan} and Teechain\cite{lind2019teechain} which leverage TEEs to establish secure, asynchronous payment channels. \textsc{A402} differs by focusing on binding off-chain service execution with on-chain asset payments. L402 applies Lightning payment channels to HTTP 402 API payments with macaroon-based authentication. 
Overall, existing payment channels are execution-unawareness and cannot prevent payment without service execution or service execution without payment.




\myparagraph{TEEs for Blockchain}
Integrating TEE with blockchain has been explored to overcome the inherent limitations of blockchains. A substantial body of work\cite{cheng2019ekiden,das2019fastkitten,huang2024advancing,kuefner2021bitcontracts,wust2020ace,li2025ethercloak,wen2025mercury,bentov2019tesseract,frassetto2023pose} utilizes TEEs to offload complex computations from the blockchain to enhance both scalability and privacy.
Unlike prior work that uses TEEs to extend on-chain computation, A402 leverages TEEs to bind off-chain service execution to on-chain assets payment.

\myparagraph{Blockchain-based Fair Exchange}
While blockchain-based fair exchange protocols have evolved from data extraction \cite{eckey2020optiswap,liu2023fairrelay,tas2024atomic,campanelli2017zero,dziembowski2018fairswap} to functional data extraction\cite{vanjani2024functional}, they remain inefficient for general-purpose programs. A402 realizes atomic exchange protocol that secures universal service execution-for-payment, extending fair exchange to arbitrary computational tasks.

\section{Conclusion}
\label{sec:conclusion}
This paper presented $\textsc{A402}$, a trust-minimized protocol that enables Web 3.0 payments for Web 2.0 services to support agentic commerce. By introducing Atomic Service Channels (ASCs), A402 employs a TEE-assisted adaptor signature scheme to enforce the Exec–Pay–Deliver atomicity for each service request. To mitigate privacy leakage, $\textsc{A402}$ incorporates a TEE-backed liquidity vault that achieves batch settlements while enabling private ASC. Evaluation on Ethereum and Bitcoin confirms that $\textsc{A402}$ achieves sub-second latency and high throughput, providing a scalable and secure payment solution for the emerging agentic commerce.

\appendix

\section{A402 on Bitcoin: Implementation and Evalution}
\label{appendix:bitcoin}
\myparagraph{Bitcoin Enforcement with Taproot}
On the Bitcoin side, we implement A402 on a Bitcoin Core v25.0 \texttt{regtest} environment.
The protocol lifecycle maps strictly to Bitcoin's UTXO model: channel creation generates a specific Pay-to-Taproot (P2TR) output (Locking)\cite{wuille2020bip341}, while channel closure consumes this output (Spending) to finalize settlement.
In standard mode, The \texttt{createASC} transaction broadcasts a transaction that instantiates a P2TR output tailored for the ASC closure. Specifically, all closure conditions are encoded within a Merkle Abstract Syntax Tree (MAST), accessible via Script-Path spending. Specifically, the MAST encodes three closure paths: (i) cooperative closure by the vault ($\mathcal{U}.\texttt{closeASC}$); (ii) force closure by the service provider, executed via adaptor signatures ($S.\texttt{forceClose}$); (iii) force closure by client, secured by relative timelocks (CSV) to allow challenge periods ($C.\texttt{forceClose}$).
In liquidity vault mode, the \texttt{initVault} transaction broadcasts a transaction that instantiates a specialized P2TR output. Similarly, all settlement conditions are encoded within a Merkle Abstract Syntax Tree (MAST), accessible via Script-Path spending. Specifically, the MAST encodes two settlement paths: (i) cooperative batch settlement by the vault ($\mathcal{U}.\texttt{forceSettle}$); (ii) force settlement by the client or service provider, secured by long relative timelocks (CSV) to serve as an emergency escape hatch during vault failure($C/S.\texttt{forceSettle})$.

\begin{table}[h]
\centering
\small
\caption{\textbf{Bitcoin On-chain Cost.} Fee Rate: 10 sat/vB, BTC Price: \$50,000. All transactions utilize a standard 1-Input, 2-Output structure, with the exception of the x402 finalization ($tx_2$) which is optimized to a single output.}
\label{tab:btc_cost}
\renewcommand{\arraystretch}{1.1}
\resizebox{\linewidth}{!}{
\begin{tabular}{lrrrr}
\toprule
\textbf{Operation} & \textbf{Base Size} & \textbf{Witness} & \textbf{Total Size} & \textbf{Cost (USD)} \\
& \textit{(vB)} & \textit{(vB)} & \textit{(vB)} & \\
\midrule
\multicolumn{5}{l}{\textbf{\textit{A402: Standard Mode (ASC)}}} \\
$\mathcal{U}$.\texttt{createASC} & 125 & 27.75 & 153 & \$0.765 \\
$\mathcal{U}$.\texttt{closeASC} & 137  & 17.25  & 155 & \$0.775 \\
\cmidrule(lr){1-5}
\textit{Total Cost($n$ request)} & \textbf{262} & \textbf{45.00} & \textbf{308} & \textbf{\$1.54} \\
\midrule
\multicolumn{5}{l}{\textit{--- Force Exit Operations ---}} \\
$S$.\texttt{forceClose} & 137 & 71.00 & 208 & \$1.04 \\
$C$.\texttt{forceClose} & 137 & 54.50 & 192 & \$0.96 \\
\midrule
\multicolumn{5}{l}{\textbf{\textit{A402: Liquidity Vault Mode}}} \\
$\mathcal{U}$.\texttt{initVault} & 125 & 27.75 & 153 & \$0.765 \\
$\mathcal{U}$.\texttt{settleVault} & 137 & 17.25 & 155 & \$0.775 \\
\cmidrule(lr){1-5}
\textit{Total Cost($n$ request)} & \textbf{262} & \textbf{45.00} & \textbf{308} & \textbf{\$1.54} \\
\midrule
\multicolumn{5}{l}{\textit{--- Force Exit Operations ---}} \\
$C/S$.\texttt{forceSettle} & 137  & 54.50 & 192 & \$0.96 \\
\midrule
\multicolumn{5}{l}{\textbf{\textit{Baseline: x402}}} \\
$tx_1$: Setup ($C \to \text{F}$) & 125 & 27.75 & 153 & \$0.765 \\
$tx_2$: Finalize ($\text{F} \to S$) & 82  & 56.00  & 138 & \$0.69 \\
\cmidrule(lr){1-5}
\textit{Total Cost($n$ request)} & \textit{207*n} & \textit{83.75*n} & \textit{291*n} & \textit{\$1.455*n} \\
\textit{100 Requests Total ($O(n)$)} & \textbf{20,700} & \textbf{8,375} & \textbf{29,100} & \textbf{\$145.50} \\
\bottomrule
\multicolumn{5}{p{0.5\textwidth}}{}
\end{tabular}}
\end{table}

\myparagraph{Transaction Cost on Bitcoin}
Table~\ref{tab:btc_cost} reports the Bitcoin on-chain transaction costs of \textsc{A402} under Taproot (P2TR) specifications.
In the standard mode, a channel lifecycle consists of two transactions: opening the ASC and closing it.
As shown in the table, $\mathcal{U}.\texttt{createASC}$ spends 153~vB (\$0.765) when funding the ASC with a P2WPKH input and creating a P2TR locking output (plus change), while the cooperative $\mathcal{U}.\texttt{closeASC}$ spends 155~vB (\$0.775) via a Taproot key-path spend with a standard 1-input, 2-output structure.
Consequently, the total on-chain cost of a complete ASC lifecycle is 308~vB (\$1.54).
In adversarial scenarios, the ASC can still be closed unilaterally via script-path spends.
A provider-initiated force closure $S.\texttt{forceClose}$ costs 208~vB (\$1.04), while a client-initiated force-close $C.\texttt{forceClose}$ costs 192~vB (\$0.96).
Relative to the cooperative close (155~vB), the additional overhead is modest (37--53~vB), and remains practical in absolute terms due to Taproot's witness discount and the fact that only a small script witness (rather than large scriptsig data) is revealed.

In liquidity vault mode, the vault is funded via $\mathcal{U}.\texttt{initVault}$ (153~vB, \$0.765) and settled via $\mathcal{U}.\texttt{settleVault}$ (155~vB, \$0.775), yielding the same constant on-chain lifecycle cost of 308~vB (\$1.54).
Importantly, a single vault settlement can aggregate payouts for many off-chain ASCs into one on-chain transaction, keeping the amortized on-chain cost per request negligible as the request volume grows.

\emph{Comparison with x402.}
To compare \textsc{A402} against x402, we model each x402 service request with two on-chain transactions.
The setup transaction ($tx_1$) locks the client's funds into an escrow UTXO (modeled as a SegWit output) and costs 153~vB (\$0.765) under the same 1-input, 2-output assumption.
The finalization transaction ($tx_2$) spends the escrow with signatures from the facilitator and the service provider and is optimized to a single output, costing 138~vB (\$0.69).
Thus, x402 incurs 291~vB (\$1.455) \emph{per request}, and its total cost scales linearly with the number of requests.

By contrast, \textsc{A402} pays a constant \$1.54 for opening and closing a channel (or funding and settling a vault), independent of how many requests are served off-chain in between.
Therefore, x402 exceeds \textsc{A402}'s fixed lifecycle cost after just two requests (\$2.91 vs.\ \$1.54), and for 100 requests, x402 costs \$145.50 whereas \textsc{A402} remains at \$1.54-- a $94\times$ reduction in on-chain fees.
This confirms that \textsc{A402} is substantially more economical for high-frequency machine-to-machine payments.

\begin{figure}[h]
    \centering
    \includegraphics[width=0.8\linewidth]{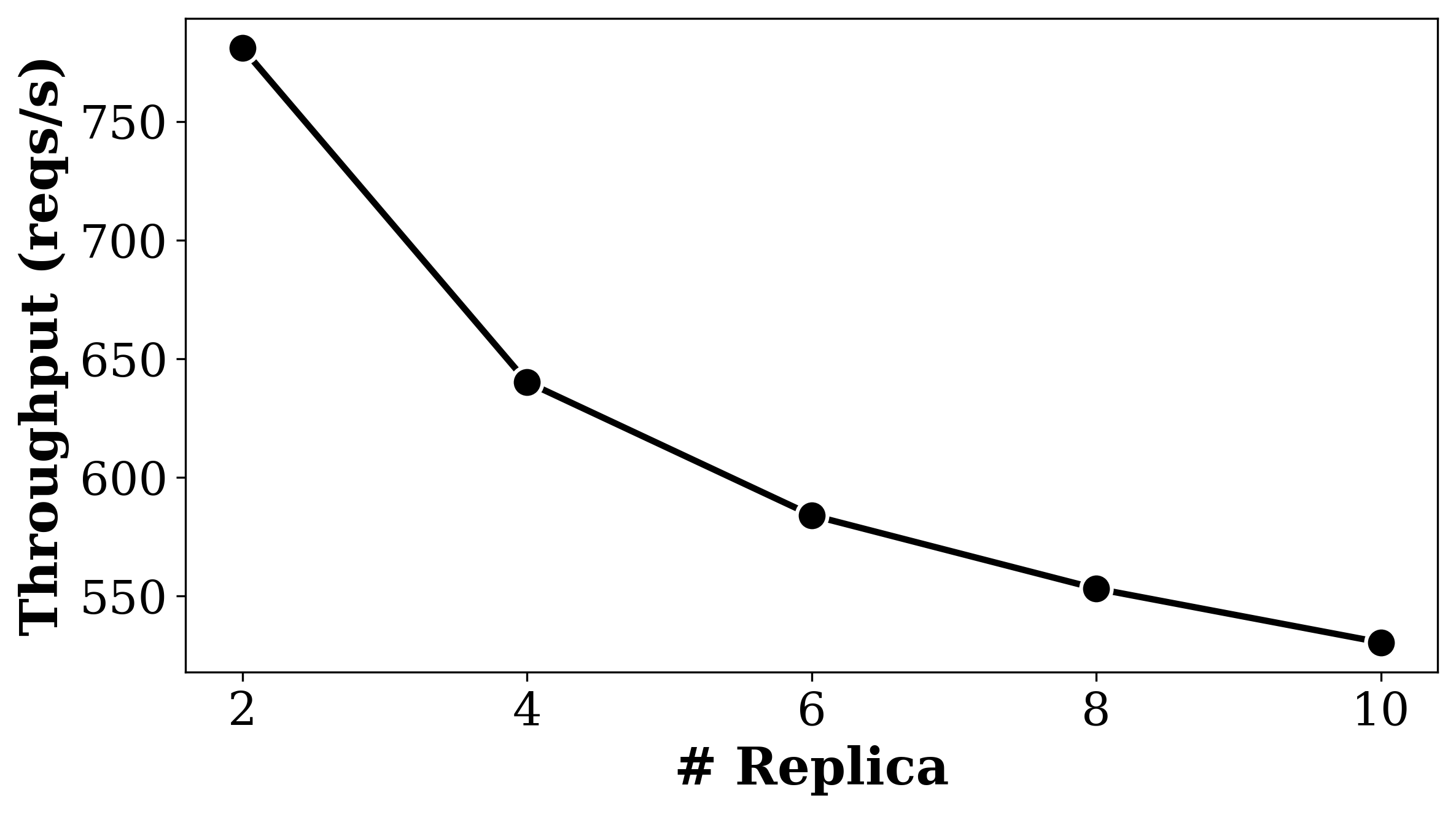}
    \caption{The throughput with varying $\mathcal{U}$ replicas (8 CPU cores per replica)}
    \label{fig:throughput2}
\end{figure}

\begin{figure}[h]
    \centering
    \includegraphics[width=0.8\linewidth]{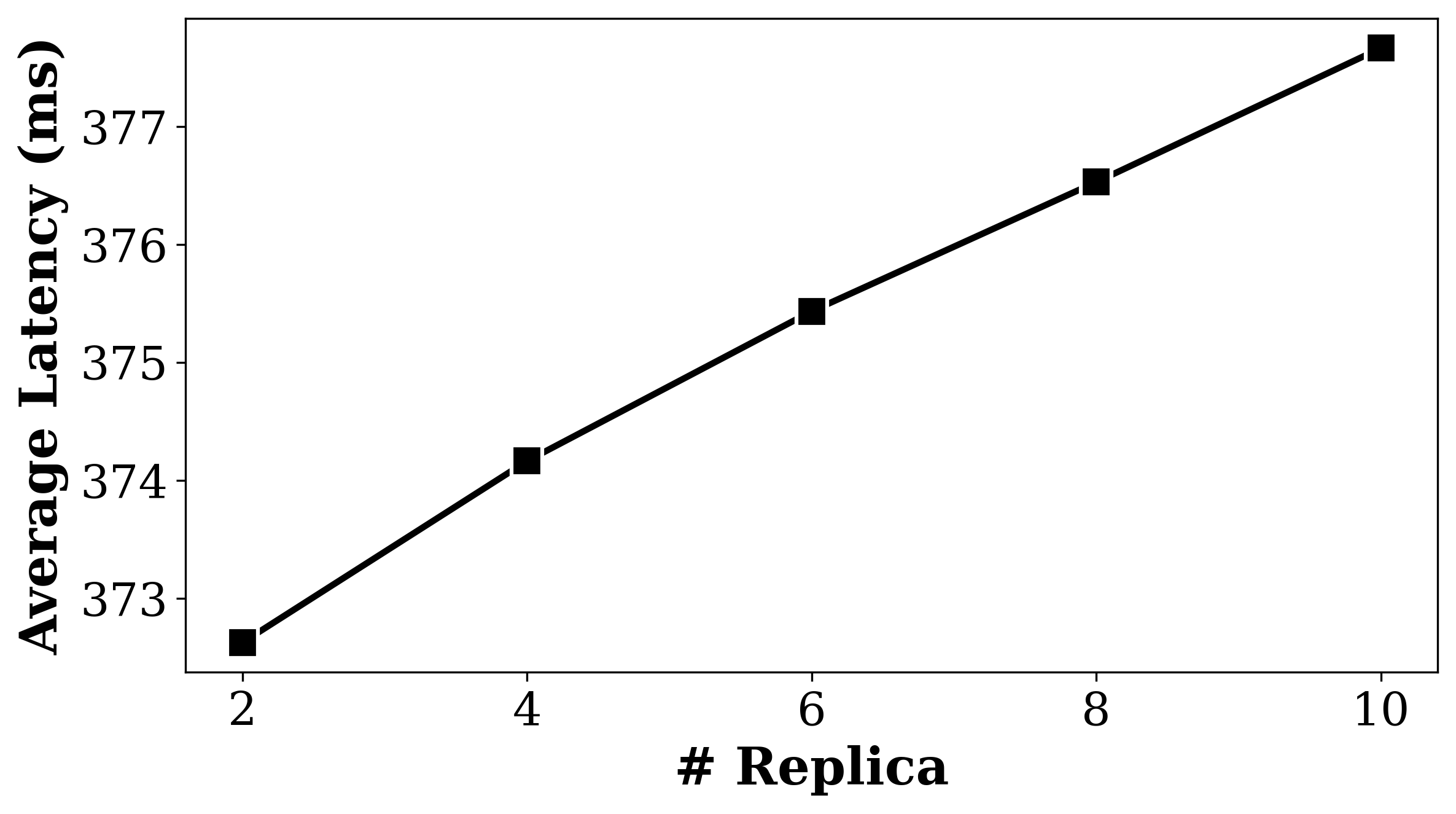}
    \caption{The latency with varying $\mathcal{U}$ replicas}
    \label{fig:latency2}
\end{figure}

\section{Scalability with Varying Replica in Vault}
\label{appendix:replica}
Compared with the single vault evaluation in Section \ref{subsec:performance}, the multi-replica experiments were conducted under a fixed request load of 1,000 requests and a fixed total CPU budget. In this setting, each replica was allocated only 8 CPU cores, whereas the single vault configuration used 64 CPU cores. As a result, the throughput of two replicas (780.86 RPS) is already significantly lower than the single-Vault throughput (2,063.14 RPS), primarily due to reduced per-replica compute capacity rather than inherent scalability limitations.

As illustrated in Figure~\ref{fig:throughput2}, throughput further decreases as the number of Vault replicas increases, dropping from 780.86 RPS with 2 replicas to 530.49 RPS with 10 replicas, while Figure~\ref{fig:latency2} shows that average latency increases only marginally from 372.63 ms to 377.66 ms. This throughput reduction is mainly attributed to additional coordination overhead introduced by replication, including signature fragment distribution, aggregation, and state synchronization across replicas, combined with reduced per-replica CPU resources. Despite this, latency remains stable (increasing by only 1.4\%) because it is dominated by simulated service execution and network delay rather than replica coordination.

\section{Summary of A402 Interfaces}
\label{appendix:interface}

For a full technical specification, we provide a detailed mapping of $\textsc{A402}$ APIs to their respective on-chain transactions in Table~\ref{tab:api_summary}. It explicitly categorizes each function by its caller ($\mathcal{U}, C, S$) and distinguishes between purely off-chain execution and those requiring on-chain transactions.

\begin{table*}[h]
\centering
\caption{Summary of \textsc{A402} System APIs and On-chain Interfaces.}
\label{tab:api_summary}
\footnotesize 
\begin{tabularx}{\textwidth}{@{} l >{\raggedright\arraybackslash}p{2.8cm} >{\raggedright\arraybackslash}p{2.5cm} >{\raggedright\arraybackslash}p{3.2cm} X @{}}
\toprule
\textbf{Protocol Stage} & \textbf{API Function} & \textbf{Bitcoin Interface} & \textbf{Ethereum Interface} & \textbf{Description} \\ \midrule
\multirow{2}{*}{\textbf{I. ASC Creation}} & $S.\mathtt{Register}$ & -- & -- & Register $S$ via remote attestation and bind identity to code hash. \\
 & $\mathcal{U}.\mathtt{reqOpenASC}$ & $\mathcal{U}.\texttt{createASC}$ & $\mathcal{U}.\texttt{createASC}$ & Initiate an ASC by locking initial assets on the underlying blockchain. \\ \midrule
\multirow{5}{*}{\textbf{II. Atomic Exchange}} & $\mathcal{U}.\mathtt{sendRequest}$ & -- & -- & Issue a request with a conditional payment. \\
 & $S.\mathtt{execRequest}$ & -- & -- & Execute the request inside the TEE and generate an encrypted response. \\
 & $\mathcal{U}.\mathtt{onExecReply}$ & -- & -- & Verify execution proof and authorize the release of conditional payment. \\
 & $S.\mathtt{revealSecret}$ & -- & -- & Reveal the witness to finalize the payment off-chain or on-chain. \\
 & $\mathcal{U}.\mathtt{onSecretReveal}$ & -- & -- & Finalize local state and deliver the decrypted result to the user. \\ \midrule
\multirow{4}{*}{\textbf{III. ASC Closure}} & $C/S.\mathtt{reqCloseASC}$ & $\mathcal{U}.\texttt{closeASC}$ & $\mathcal{U}.\texttt{closeASC}$ & Cooperative closure requested to $\mathcal{U}$ for immediate settlement. \\
 & $C.\mathtt{forceCloseASC}$ & $C.\texttt{forceClose}$ & $C.\texttt{initForceClose}$ \newline $C.\texttt{finalForceClose}$ & Unilateral closure by $C$; requires a dispute window on Ethereum. \\
 & $S.\mathtt{forceCloseASC}$ & $S.\texttt{forceClose}$ & $S.\texttt{forceClose}$ & Unilateral closure by $S$ via adaptor signature; no challenge period. \\ \midrule
\multirow{5}{*}{\textbf{Liquidity Vault}} & $\mathcal{U}.\mathtt{initVault}$ & $\mathcal{U}.\texttt{initVault}$ & $\mathcal{U}.\texttt{initVault}$ & Deposit assets into the TEE-backed vault to create liquidity vault. \\
 & $\mathcal{U}.\mathtt{reqOpenVaultASC}$ & -- & -- & Create a private ASC and bind specified vault assets to ASC. \\
 & $C/S.\mathtt{reqCloseVaultASC}$ & -- & -- & Off-chain closure of a private ASC via $\mathcal{U}$ without on-chain transactions. \\
 & $\mathcal{U}.\mathtt{settleVault}$ & $\mathcal{U}.\texttt{settleVault}$ & $\mathcal{U}.\texttt{settleVault}$ & Aggregate multiple private ASC settlement for a on-chain withdrawal. \\
 & $C/S.\mathtt{forceSettleVault}$ & $C/S.\texttt{forceSettle}$ & $C/S.\texttt{initForceSettle}$ \newline $C/S.\texttt{finalForceSettle}$ & Emergency exit from vault; requires dispute window on blockchain. \\ \bottomrule
\end{tabularx}
\end{table*}

{\footnotesize \bibliographystyle{acm}
\bibliography{sample}}

\end{document}